\documentstyle[tighten,preprint,eqsecnum,aps,prd]{revtex}
\input epsf

\begin{document}
\draft

\def\Bbb{\bf}
\def\BbbR{{\Bbb R}}
\def\BbbZ{{\Bbb Z}}
\def\d{\Delta}
\def\RPthree{{\BbbR P^3}}
\def\arsinh{\mathop{\rm arsinh}\nolimits}

\preprint{\vbox{\baselineskip=12pt
\rightline{WISC--MILW--97--TH--10}
\rightline{IUCAA--7/97}
\rightline{gr-qc/9702038}}}
\title{Lorentzian Approach to Black Hole Thermodynamics in the 
       Hamiltonian Formulation}
\author{Sukanta Bose,\footnote{Electronic address:
{\em sbose@iucaa.ernet.in}}
}
\address{IUCAA, Post Bag 4, Ganeshkhind, Pune 411007, India}
\author{
Leonard Parker\footnote{Electronic address:
{\em leonard@cosmos.phys.uwm.edu}}
and,
Yoav Peleg\footnote{Electronic address:
{\em yoavp@msil.sps.mot.com}}
}
\address{Department of Physics,
University of Wisconsin-Milwaukee, \\
P.~O. Box 413,
 Milwaukee, Wisconsin 53201, USA}
\date{February 1997}

\maketitle
\begin{abstract}%

In this work, we extend the analysis of Brown and York to find the 
quasilocal energy in a spherical box in the Schwarzschild spacetime. 
Quasilocal energy is the value of the Hamiltonian that generates
unit magnitude proper-time translations on the box orthogonal to the
spatial hypersurfaces foliating the Schwarzschild spacetime.
We call this Hamiltonian the Brown-York Hamiltonian.
We find different classes of foliations that correspond to time-evolution 
by the Brown-York Hamiltonian. We show that although the Brown-York 
expression for the quasilocal energy is correct, one needs to supplement 
their derivation with an extra set of boundary conditions on the interior 
end of the spatial hypersurfaces inside the hole in order to obtain it
from an action principle. Replacing this set of boundary conditions with 
another set yields the Louko-Whiting Hamiltonian, which corresponds to
time-evolution of spatial hypersurfaces in a different foliation of
the Schwarzschild spacetime.
We argue that in the thermodynamical picture, the Brown-York Hamiltonian
corresponds to the {\em internal energy} whereas the Louko-Whiting 
Hamiltonian corresponds to the {\em Helmholtz free energy} of the 
system. Unlike what has been the usual route to black hole
thermodynamics in the past, this observation immediately allows us to 
obtain the partition function of such a system without resorting to any 
kind of Euclideanization of either the Hamiltonian or the action. In the 
process, we obtain some interesting insights into the geometrical nature
of black hole thermodynamics.

\end{abstract}

\narrowtext

\section{Introduction}
\label{sec:intro}

After more than two decades of investigations, black hole 
thermodynamics is still one of the most puzzling subjects in 
theoretical physics. One approach to studying the thermodynamical
aspects of a black hole involves considering the evolution of quantum matter 
fields propagating on a classical (curved) background spacetime. 
This gives rise to the phenomenon of black hole radiation that was 
discovered by Hawking in 1974 \cite{Hawking1}.
Combining Hawking's discovery of black hole radiance 
with the classical laws of black hole mechanics \cite{Carter}, 
leads to the laws of black hole thermodynamics. The entropy of a black
hole obtained from this approach may be interpreted as resulting from 
averaging over the matter field degrees of freedom lying  
either inside the black hole \cite{BKLS} or, equivalently, outside the
black hole \cite{FN}, as was first anticipated by Bekenstein 
\cite{Bekenstein} even before Hawking's discovery. The above approach 
was further developed in the following years \cite{Birrell,Srednicki}. 

A second route to black hole thermodynamics involves using the 
path-integral approach to quantum gravity to study {\em vacuum} 
spacetimes (i.e., spacetimes without matter fields). In this method,
the thermodynamical partition function is computed from the propagator
in the saddle point approximation \cite{GH1,hawkingCC} and it leads to 
the same laws of black hole thermodynamics as obtained by the 
first method. 
The second approach was further developed in the following years 
\cite{york1,WYprl,whitingCQG,pagerev,LWo,BY-microcan,MW}. 
The fact that the laws of black hole thermodynamics can be derived 
without considering matter fields, suggests that there may be a purely  
geometrical (spacetime) origin of these laws. However, a complete 
geometrical understanding of black hole thermodynamics is not yet present. 

In general, a basic understanding of the thermodynamical properties 
of a system requires a specification of the system's (dynamical) 
degrees of freedom (d.o.f.). Obtaining such a specification is a nontrivial 
matter in quantum gravity. In the path-integral approach one avoids the 
discussion of the dynamical d.o.f.. There, the dominant contribution to the
partition function comes from a saddle point, which is a classical 
Euclidean solution \cite{GH1}. Calculating the contribution of such a 
solution to the partition function does not require an identification of 
what the dynamical d.o.f.'s of this solution are.
Though providing us with an elegant way of getting the laws of 
black hole thermodynamics, the path-integral approach does not give us 
the basic (dynamical) d.o.f. from which we can have a better geometrical 
understanding of the origin of black hole thermodynamics. 

It was only recently that the dynamical geometric d.o.f. for a spherically 
symmetric vacuum Schwarzschild black hole were found \cite{TK,KVK} under
certain boundary conditions \footnote{We thank Jorma Louko for suggesting
Refs. \cite{TK} to us.}. In particular, by considering general foliations 
of the complete Kruskal extension of the Schrawzschild spacetime, Kucha\v{r} 
\cite{KVK} finds a reduced system of {\em one} pair of canonical variables 
that can be viewed as global geometric d.o.f.. One of these is the 
Schwarzschild mass, while the other one, its conjugate momentum, is the 
difference between the parametrization times at right and left spatial
infinities. Using the approach of Kucha\v{r}, recently Louko 
and Whiting \cite{LW} (henceforth referred to as LW) studied black hole
thermodynamics in the Hamiltonian formulation. As shown in Fig. 2, they 
considered a foliation in which the spatial hypersurfaces are restricted to 
lie in the right exterior region of the Kruskal diagram and found 
the corresponding reduced phase space system. This enabled them to 
find the unconstrained Hamiltonian (which evolves these spatial hypersurfaces)
and canonically quantize this reduced theory. They then obtain the
Schr\"{o}dinger time-evolution operator in terms of the reduced 
Hamiltonian. The partition function $Z$ is defined as 
the trace of the Euclideanised time-evolution operator $\hat{K}_{\rm E}$,
namely, $Z = \mbox{Tr}(\hat{K}_{\rm E})$, where the hat denotes a quantum 
operator. This partition function has the same 
expression as the one obtained from the path-integral approach and 
expectedly yields the laws of black hole thermodynamics. 

In a standard thermodynamical system it is not essential to consider 
{\em Euclidean}-time action in order to study the thermodynamics. 
If $\hat{H}$ 
is the Lorentzian time-independent Hamiltonian of the system, 
then the partition function is defined as 
\begin{equation}
\label{partition-trace}
Z = \mbox{Tr exp}(- \beta \hat{H})
\ \ ,
\end{equation} 
where $\beta$ is the inverse temperature 
of the system in equilibrium. However, in many cases (especially, in time-
independent systems) the Euclidean time-evolution operator turns out to
be the same as $\exp (- \beta \hat{H})$. Nevertheless, there are cases where,
as we will see in section \ref{subsec:LWHam}, 
the Euclidean time-evolution operator is not the same as 
$\mbox{exp}(-\beta \hat{H})$. This is the case 
for example in the LW approach, i.e., $(\hat{K}_{\rm E})_{\rm LW} \neq 
\mbox{exp}(-\beta \hat{\sf h})$, where $\hat{\sf h}$ is the reduced 
Hamiltonian 
of the quantized LW system. There is a geometrical reason for this 
inequality and in this work we discuss it in detail. In this paper, 
we ask if there exists a Hamiltonian $\hat{H}$ (which is associated with 
certain foliations of the Schwarzschild spacetime) appropriate for
finding the partition function of a Schwarzschild black hole enclosed 
inside a finite-sized box using (\ref{partition-trace}). 
Such a procedure will not resort to Euclideanization. In our quest to 
obtain the Hamiltonian that is appropriate for defining the partition
function for (\ref{partition-trace}), we also clarify the physical
significance of the LW Hamiltonian. By doing so we hope to achieve a 
better understanding of the geometrical origin of the
thermodynamical aspects of a black hole spacetime.

In a previous work \cite{BY-quasilocal}, Brown and York (henceforth 
referred to as BY) found a general expression for the quasilocal energy 
on a timelike two-surface that bounds a spatial three-surface located in a 
spacetime region that can be decomposed as a product of a spatial 
three-surface and
a real line interval representing time. From this expression they obtained
the quasilocal energy inside a spherical box centered at the origin of 
a four-dimensional 
spherically symmetric spacetime. They argued that this expression also 
gives the correct quasilocal energy on a box in the Schwarzschild spacetime.
In this paper we show that, although their expression for the 
quasilocal energy on a box in the Schwarzschild spacetime is correct,
the analysis they use to obtain it requires to be extended
when applied to the case of Schwarzschild spacetime. In this case, one
needs to impose extra boundary conditions at the timelike boundary 
inside the hole (see Fig. 3). As mentioned above, in principle, one can use 
the Hamiltonian $\hat{\sf H}$ so obtained to evaluate the
partition function, $Z={\rm Tr}\exp(-\beta \hat{\sf H})$. This partition
function corresponds to the canonical ensemble and describes the 
thermodynamics of a system whose volume and temperature are fixed but 
whose energy content is permitted to vary. Such a Hamiltonian, 
$\hat{\sf H}$ would then lead to a description of black hole 
thermodynamics without any sort of Euclideanisation. The only obstacle
to this route to the partition function is that the trace can be evaluated 
only if one knows the
density of the energy eigenstates. Unfortunately, without knowing what 
the thermodynamical entropy of the system is, it is not clear how 
to find this density in terms of the reduced phase-space variables of
Kucha\v{r} \cite{KVK}. So how can one derive the thermodynamical laws 
of the Schwarzschild black hole using a Lorentzian Hamiltonian without 
knowing the density of states?

Based on an observation that identifies the thermodynamical roles of the
BY and the LW Hamiltonians we succeed in studying black hole thermodynamics
within the Hamiltonian formulation but without Euclideanization.
In section \ref{sec:thermoconsi} we describe the thermodynamical roles of the
BY and the LW Hamiltonians. Identifying these roles allows us to immediately
calculate the partition function and recover the thermodynamical properties
of the Schwarzschild black hole. In section \ref{sec:geoconsi} we study the
geometrical significance of these Hamiltonians. In particular, we extend the
work of Brown and York \cite{BY-quasilocal} to find the nature of the spatial
slices that are evolved by the BY Hamiltonian in the full Kruskal extension 
of the Schwarzschild spacetime. In section \ref{geothermo} we use the 
observations made in sections \ref{sec:thermoconsi} and \ref{sec:geoconsi}
to ascribe geometrical 
basis to the thermodynamical parameters of the system, thus gaining 
insight into the geometrical nature of black hole thermodynamics.

We conclude the paper in section \ref{sec:conclu} by summarising our results
and discussing the connection between the foliation geometry and 
equilibrium black hole thermodynamics. In appendix A we extend our 
results to the case of two-dimensional dilatonic black holes. In appendix B
we discuss an alternative foliation (see Fig. 4), in which the spatial slices
are again evolved by the BY Hamiltonian ${\sf H}$. This illustrates the 
non-uniqueness of the foliation associated with the BY Hamiltonian.

We shall work throughout in ``geometrized-units'' in which $c=G=1$. 

\section{Thermodynamical Considerations}
\label{sec:thermoconsi}

\subsection{The Brown-York Hamiltonian}
\label{subsec:BYHam}

It was shown by Brown and York \cite{BY-quasilocal} that
in 4D spherically symmetric Einstein gravity, the quasilocal energy of 
a system that is enclosed
inside a spherical box of finite surface area and which can be
embedded in an asymptotically flat space is \cite{BY-quasilocal}
\begin{equation}
\label{quasiH}
{\sf H} = \left( 1 -\sqrt{1-{2{\bf m} \over B}}\right) B,
\end{equation}
where ${\bf m}$ is the ADM mass of the spacetime and $B$ is the
fixed curvature radius of the box with its origin at the center of
symmetry. We will call ${\sf H}$ the Brown-York Hamiltonian.
 
The Brown and York derivation of the quasilocal energy can be summarised
as follows. The system they consider is a spatial three-surface $\Sigma$
bounded by a two-surface ${\sf B}$ in a spacetime region that can be 
decomposed as a product of a spatial three-surface and a real line 
interval representing time (see Fig. 1). The time evolution of the 
two-surface boundary ${\sf B}$ is the timelike three-surface boundary 
${}^3 {\sf B}$.  They then obtain a surface stress-tensor on the boundary by
taking the functional derivative of the action with respect to the
three-metric on ${}^3 {\sf B}$. The energy surface density is the
projection of the surface stress tensor normal to a family of 
spacelike two-surfaces like ${\sf B}$ that foliate ${}^3 {\sf B}$.
The integral of the energy surface density over such a two-surface
${\sf B}$ is the quasilocal energy associated with a spacelike 
three-surface $\Sigma$ whose {\em orthogonal} ~intersection with 
${}^3 {\sf B}$ is the two-boundary ${\sf B}$. 

As argued by BY, Eq. (\ref{quasiH}) also describes the total energy 
content of a box enclosing a Schwarzschild hole. One would thus expect to
obtain the corresponding partition function from it by the prescription 
$Z={\rm Tr}\exp(-\beta \hat{\sf H})$. As mentioned above, the only 
obstacle to this calculation is the lack of knowledge about the 
density of states of the system, which is needed to evaluate the trace.
However, as we discuss in the next subsection, there is another Hamiltonian
associated with the Schwarzschild spacetime that allows us to obtain
the relevant partition function without Euclideanization. This is the 
Louko-Whiting Hamiltonian.

\subsection{The Louko-Whiting Hamiltonian}
\label{subsec:LWHam}

In their quest to obtain the partition function for the Schwarzschild 
black hole in the Hamiltonian formulation, LW found a Hamiltonian that
time-evolves spatial hypersurfaces in a Schwarzschild spacetime of
mass ${\bf m}$ such that 
the hypersurfaces extend from the bifurcation 2-sphere to a timelike 
box-trajectory placed at a constant curvature radius of $R=B$ 
(see Fig. 2). As we will show in the next subsection,
the LW Hamiltonian describes the correct free energy of a Schwarzschild
black hole enclosed inside a box in the thermodynamical picture. The LW 
Hamiltonian is
\begin{equation}
\label{LWht}
{\bf h}(t) = \left( 1 -\sqrt{1-{2 {\bf m} \over B}}\right) B Q_B(t) - 
          2{\bf N}_0 (t) {\bf m}^2, 
\end{equation}
where generically $Q_B$ and ${\bf N}_0$ are functions of time $t$,
which labels the spatial hypersurfaces. 
Physically, $Q_B\equiv \sqrt{-g_{tt}}\>$, where $(-g_{tt})$ is 
the time-time component of the spacetime metric on the box.  
On the other hand, the physical meaning of ${\bf N}_0$
is as follows. On a classical solution, consider the future timelike 
unit normal $n^a (t)$ to a constant $t$ hypersurface at the bifurcation 
two-sphere (see Fig. 2). Then ${\bf N}_0$ is the rate at which the constant 
$t$ hypersurfaces are boosted at the bifurcation 2-sphere:
\begin{equation}
n^a (t_1) n_a (t_2) = - \cosh \left( \int_{t_1}^{t_2} {\bf N}_0 \> dt
                              \right)
\ \ ,
\end{equation}
where $t=t_1$ is the initial hypersurface and $t=t_2$ is the boosted
hypersurface.

If one is restricted to a foliation in which the spatial
hypersurfaces approach the box along surfaces of constant proper time
on the box, 
then $Q_B=1$. On classical solutions, the spatial hypersurfaces approach the
bifurcation 2-sphere along constant Killing-time hypersurfaces (see 
the paragraph containing Eqs. (\ref{ls-r}) in section \ref{subsec:LWgeoham}). 
The LW fall-off conditions (\ref{ls-r}), 
which are imposed on the ADM variables at the bifurcation 2-sphere, can be
used to show that on solutions, ${\bf N}_0 = \kappa \> dT / dt$, where $T$ is 
the Killing time and $\kappa = (4{\bf m})^{-1}$ is the 
surface gravity of a Schwarzschild black hole.
In the particular case where the label time $t$ is taken to be the 
proper time on the box, we have
\begin{equation}
\label{N0sol}
{\bf N}_0 = \left[ 4 {\bf m} \sqrt{1 - {2 {\bf m} \over B}} \> \right]^{-1}
\ \ .
\end{equation}
With such an identification of the label time $t$, Eq. (\ref{LWht})
shows that on classical solutions the time-evolution of these spatial 
hypersurfaces is given by the Hamiltonian 
\begin{equation}
\label{LWh}
{\sf h} = \left( 1 -\sqrt{1-{2{\bf m}\over B}}\right) B  - 
          2{\bf N}_0  {\bf m}^2, 
\end{equation}
where ${\bf N}_0$ is given by (\ref{N0sol}).

One may now ask if one can use the LW Hamiltonian (\ref{LWh})
to obtain a partition function for the system and study its thermodynamical
properties. Unfortunately, one cannot do so in a straightforward 
manner. First, one cannot replace $\hat{H}$ in (\ref{partition-trace}) 
by $\hat{\sf h}$, the quantum counterpart of (\ref{LWh}), to obtain the 
partition function. The reason is that classically ${\sf h}$ does not 
give the correct energy of the system; the BY Hamiltonian ${\sf H}$ of
(\ref{quasiH}) does. To avoid this problem, LW first construct the 
Schr\"{o}dinger time-evolution operator $\hat{K}\equiv \exp (-i\int 
\hat{\sf h} dt)$. They then Euclideanize this operator and use it to obtain 
the partition function $Z = {\rm Tr} ({\hat{K}}_{\rm E})$. The partition
function so obtained does not equal ${\rm Tr} \exp (-\beta \hat{\sf h})$,
but rather it turns out to be the same as that obtained via the
path integral approach of Gibbons and Hawking \cite{GH1}. However, apart 
from this
end result, a justification at some fundamental level has been lacking 
as to why the LW Hamiltonian (\ref{LWh}) and not any other Hamiltonian 
(eg., (\ref{quasiH})) should be used to obtain the partition function 
using the LW procedure. 

In the next subsection, we will find the thermodynamical roles played by the 
BY and LW Hamiltonians. We will also show how this helps us in obtaining the
partition function without Euclideanization. This way we will avoid the
ambiguity mentioned above that arises in the LW-method of constructing the
partition function.

\subsection{Internal Energy and Free Energy}
\label{subsec:free-energy}

As argued by Brown and York \cite{BY-quasilocal}, on solutions,
the BY Hamiltonian ${\sf H}$ in Eq. (\ref{quasiH}) denotes the internal
energy ${\cal E}$ residing within the box:
\begin{equation}
\label{E}
{\cal E} = \left( 1 -\sqrt{1-{2 {\bf m} \over B}}\right) B.
\end{equation}
In fact Eq. (\ref{E}) can be shown to yield the
first law of black hole thermodynamics 
\begin{equation}
\label{dE}
\d{\cal E} = -s\>\d(4\pi B^2) + \left( 8\pi {\bf m} \sqrt{1-{2 {\bf m} \over B}}
     \right)^{-1}\d(4\pi {\bf m}^2), 
\end{equation}
where $s$ is the surface pressure on the box-wall
\cite{BY-quasilocal}
\begin{equation}
\label{surfacepr}
s\equiv {1\over 8\pi B}\left( {1-{\bf m}/B \over \sqrt{1-2{\bf m}/B}}-1\right).
\end{equation}
The first term on the rhs of (\ref{dE}) is negative of the amount
of work done by the system on its surroundings and, with hindsight,
the second term is the product ${\cal T}\d{\cal S}$, where ${\cal T}$
is the temperature and ${\cal S}$ is the entropy of the system. We will 
not assume the latter in
the following analysis; rather we will deduce the form of ${\cal T}$ and 
${\cal S}$ from first principles. 

We next show that the LW Hamiltonian ${\sf h}$ of Eq. (\ref{LWh}) plays 
the role of Helmholtz free energy of the system. Recall that the Helmholtz 
free energy ${\cal F}$ is defined as
\begin{equation}
\label{F}
{\cal F} =\> {\cal E}\> -\>{\cal T}{\cal S},
\end{equation} 
where ${\cal E}$ is the internal energy. Thus in an isothermal and 
reversible process, the first law of thermodynamics implies that
the amount of mechanical work done by a system, $W$, is equal to
the decrease in its free energy, i.e.,
\begin{equation} 
\label{FW}
W = -\d{\cal F}
\ \ .
\end{equation}
As a corollary to this statement
it follows that for a mechanically isolated system at a constant
temperature, the state of equilibrium is the state of minimum free
energy. 

We now show that under certain conditions on the foliation of the
spacetime with spatial hypersurfaces, the LW Hamiltonian 
${\sf h}$ in Eq. (\ref{LWh}) plays the role of {\em free energy}. We 
choose a foliation such that on solutions the lapse ${\bf N}_0$
obeys (\ref{N0sol}). Using the expression (\ref{E}) for ${\cal E}$, 
the Hamiltonian ${\sf h}$ in Eq. (\ref{LWh}) can be rewritten as 
\begin{equation}
\label{hitoE}
{\sf h} = {\cal E} - 2 {\bf N}_0 {\bf m}^2
\ \ .
\end{equation}
Now let us perturb ${\sf h}$ about a solution by perturbing 
${\bf m}$ and $B$ such that ${\bf N}_0$ itself is held fixed. Then 
\begin{equation}
\label{dhitodE}
\d{\sf h} = \d{\cal E}- {{\bf N}_0 \over 2\pi} \d(4 \pi {\bf m}^2)
\ \ .
\end{equation}
Note that keeping ${\bf N}_0$ fixed, i.e., $\d{\bf N}_0 = 0$,
does not necessarily imply through (\ref{N0sol}) that $\d{\bf m}$ and 
$\d B$ are not independent perturbations. This is because, in general, the
perturbed ${\sf h}$ may not correspond to a solution and hence the 
perturbed ${\bf N}_0$ need not have the form (\ref{N0sol}). However, here
we will assume that the perturbations do not take us off the space of
static solutions
and, therefore, the perturbed ${\bf N}_0$ has the form (\ref{N0sol}).
Hence in our case $\d {\bf m}$ and $\d B$ are not independent perturbations.
Using (\ref{dE}) and (\ref{N0sol}) in (\ref{dhitodE}) yields
\begin{equation}
\label{dhdW}
\d{\sf h} = - s \d (4\pi B^2) = - W
\ \ .
\end{equation}
Finally, from (\ref{dhdW}) and (\ref{FW}) we get
\begin{equation}
\label{hFc}
{\sf h} = {\cal F} + c
\ \ ,
\end{equation}
where $c$ is a constant independent of ${\bf m}$. To find $c$, we take the 
limit ${\bf m} \to 0$. In this limit both ${\sf h}$ and ${\cal F}$ vanish 
and, therefore, $c$ has to be zero.

Another way to see that $c$ should vanish is to identify the geometric
quantity ${\bf N}_0$ with the temperature, ${\cal T}$, of the 
system (up to a multiplicative constant). Then the perturbation 
(\ref{dhitodE}) in ${\sf h}$, keeping ${\bf N}_0$ (and, therefore, 
${\cal T}$) fixed, describes an isothermal process. But Eq. (\ref{hFc})
shows that $c$ has to be an extensive function of thermodynamic invariants 
of the isothermal process since ${\sf h}$ and ${\cal F}$ are both extensive. 
The only thermodynamic quantity that we assume to be invariant in this 
isothermal process is the temperature ${\cal T}$. But since ${\cal T}$ is 
not extensive, $c$ has to be zero. The fact that ${\bf N}_0$ indeed 
determines the temperature of the system will be discussed in detail in a 
later section.

The above proof of the LW Hamiltonian being the Helmholtz free energy 
immediately allows us to calculate the partition function for a canonical 
ensemble of such systems,
\begin{equation}
\label{partitionh}
Z = \exp \left( -\beta {\cal F} \right)
\ \ ,
\end{equation}
by simply putting ${\cal F}={\sf h}$. In this way we 
recover the thermodynamical properties of a Schwarzschild black hole
without Euclideanization.
We will do so in detail in section \ref{geothermo} but first we establish 
the geometrical significance of BY and LW Hamiltonians in the next section.

\section{Geometrical considerations: Dynamics}
\label{sec:geoconsi}

A study of the geometrical roles of the BY and LW 
Hamiltonians provides the geometrical basis for the 
thermodynamical parameters associated with a black hole that were 
discussed in the preceeding section. In this section we begin by setting up 
the Hamiltonian formulation
appropriate for the two sets of boundary conditions that lead to the BY 
and LW Hamiltonians as being the unconstrained Hamiltonians that generate
time-evolution of foliations in Schwarzschild spacetime. The notation 
follows that of Kucha\v{r} \cite{KVK} and LW.

A general spherically symmetric spacetime
metric on the manifold $\BbbR \times \BbbR \times S^2$ can be
written in the ADM form as
\begin{equation}
ds^2 = - N^2 dt^2 + \Lambda^2 {(dr + N^r dt)}^2 +R^2 d\Omega^2
\ \ ,
\label{4-metric}
\end{equation}
where $N$, $N^r$, $\Lambda$ and $R$ are functions of $t$ and $r$ only,
and $d\Omega^2$ is the metric on the unit two-sphere. We will choose
our boundary conditions in such a way that the radial proper
distance $\int \Lambda \, dr$ on the constant $t$ surfaces is finite.
This implies that the radial coordinate $r$ have a finite range, which
we take to be $[0,1]$, without any loss of generality.
The spatial metric and the spacetime metric will be 
assumed to be nondegenerate, in particular, $\Lambda$, $R$, and $N$ 
are taken to be positive.

For the metric (\ref{4-metric}), the Einstein-Hilbert action is
\begin{eqnarray}
S_\Sigma [R, \Lambda ; N, N^r] &&
\nonumber
\\
= \int dt \int_0^1 d r \,
\bigg[
&&
-N^{-1}
\left(
R \bigl( - {\dot \Lambda}
+ (\Lambda N^r)' \bigr)
( - {\dot R} + R' N^r )
+ \case{1}{2} \Lambda
{( - {\dot R} + R' N^r )}^2
\right)
\nonumber
\\
&&
+ N
\left(
- \Lambda^{-1} R R''
+ \Lambda^{-2} R R' \Lambda'
- \case{1}{2} \Lambda^{-1} {R'}^{2}
+ \case{1}{2} \Lambda \right) \bigg]
\ \ ,
\label{S-lag}
\end{eqnarray}
where the subscript $\Sigma$ denotes that $S_{\Sigma}$ is a hypersurface 
action that is defined only up to the possible addition of boundary terms. 
Above, the overdot and the prime denote
$\case {\partial}{\partial t}$ ~and~ $\case {\partial}{\partial r}$,
respectively. The equations of motion derived from (\ref{S-lag}) are the 
full Einstein equations for the metric~(\ref{4-metric}), and they imply
that every classical solution is part of a maximally extended
Schwarzschild spacetime, where the value of the Schwarzschild mass $M$
may be positive, negative, or zero. In what follows, we will choose our
boundary conditions such that $M>0$. We shall discuss the boundary
conditions and the boundary terms after passing to the Hamiltonian
formulation.

The momenta conjugate to $\Lambda$ and $R$ are found 
from the Lagrangian action~(\ref{S-lag}) to be
\begin{mathletters}
\begin{eqnarray}
P_{\Lambda} &=& - N^{-1} R \left( {\dot R} - R' N^r \right)
\ \ ,
\label{PLambda}
\\
P_R &=& -N^{-1}
\left( \Lambda
( {\dot R} - R' N^r )
+ R \bigl(
{\dot \Lambda} - (\Lambda N^r)'
\bigr)
\right)
\ \ .
\label{PR}
\end{eqnarray}
\end{mathletters}%
A dual-Legendre transformation then yields the Hamiltonian action
\begin{equation}
\label{canhypaction}
S_\Sigma [\Lambda, R, P_\Lambda, P_R ; N, N^r]
= \int dt
\int_0^1 dr \left( P_\Lambda {\dot \Lambda} +
P_R {\dot R} - NH - N^r H_r \right)
\ \ ,
\label{S-ham}
\end{equation}
where the super-Hamiltonian $H$
and the radial supermomentum $H_r$
are 
\begin{mathletters}
\begin{eqnarray}
H &=& - R^{-1} P_R P_\Lambda
+ \case{1}{2} R^{-2} \Lambda P_\Lambda^2
+ \Lambda^{-1} R R'' - \Lambda^{-2} R R' \Lambda'
+ \case{1}{2} \Lambda ^{-1} {R'}^2
- \case{1}{2} \Lambda
\ \ ,
\label{superham}
\\
H_r &=& P_R R' - \Lambda P_\Lambda'
\ \ .
\label{supermom}
\end{eqnarray}
\end{mathletters}%
It can be verified that 
the Poisson brackets of the constraints close according to the radial
version of the Dirac algebra \cite{teitel-dirac}.

We next consider the boundary terms that must be added to the hypersurface 
action (\ref{canhypaction}) for the total action to yield, upon variation, 
only a volume term corresponding to the equations of motion. However, the 
boundary terms depend intricately on the choice of the spacetime foliation.
Different foliations require different boundary conditions on the geometric
variables in the variational analysis, thus requiring the addition of 
different boundary terms to (\ref{canhypaction}). As is well known in general 
relativity, it is these boundary terms that determine the true Hamiltonian of
the system. Hence, as we show below explicitly, this implies 
that different foliations correspond to different Hamiltonians, which are
the generators of time-evolution of the spatial slices in the foliations.

\subsection{The Brown-York Hamiltonian}
\label{subsec:BYgeoham}

In general, the analysis of Brown and York (see subsection 
\ref{subsec:BYHam}) breaks down in 
cases where spacetime regions of non-trivial topologies are
enclosed inside the spherical box, particularly so in the case of a
Schwarzschild black hole enclosed inside the box. As we show below,
in this case one is
forced to introduce an ``inner''-boundary where the spatial
hypersurfaces of Brown and York must extend to. This fact becomes
transparent when one looks at the full Kruskal extension of the
Schwarzschild metric. There, one begins by performing a $(3+1)$
decomposition of the spacetime in terms of a one-parameter family of
spatial hypersurfaces. The Kruskal diagram (see Figs. 2 and 3)
shows that any such
foliation would necessarily require two boundaries: an outer
boundary and an inner boundary. The Hamiltonian 
that evolves these
spatial hypersurfaces in time will in general depend on the boundary
conditions specified on these 2-boundaries. In this section we show
that the spatial hypersurfaces that are evolved by the Hamiltonian
corresponding to the quasilocal energy, given in (\ref{quasiH}), are
ones that extend from the box (the outer timelike boundary),
on the right end, to an inner timelike boundary located completely 
inside the dynamical region of the Kruskal diagram, on the left end. 

\subsubsection{Boundary conditions}
\label{subsec:BYgeo}

We first find the boundary conditions and the foliation that correspond to 
time evolution by the BY Hamiltonian and later we compare these with those
corresponding to the LW Hamiltonian. At $r=0$, we prescribe the 
following fall-off conditions 
\begin{mathletters}
\label{s-r}
\begin{eqnarray}
\Lambda (t,r) &=& \Lambda_0(t) + O(r^2)
\ \ ,
\label{s-r-Lambda}
\\
R(t,r) &=& R_0(t) + R_2(t) r^2 + O(r^4)
\ \ ,
\label{s-r-R}
\\
P_{\Lambda}(t,r) &=& P_{\Lambda 0}(t) + O(r^2)
\ \ ,
\label{s-r-PLambda}
\\
P_{R}(t,r) &=& P_{R 0}(t) + O(r^2)
\ \ ,
\label{s-r-PR}
\\
N(t,r) &=& N_0(t) + O(r^2)
\ \ ,
\label{s-r-N}
\\
N^r(t,r) &=& N^r_1(t)r + O(r^3)
\ \ ,
\label{s-r-Nr}
\end{eqnarray}
\end{mathletters}%
where $\Lambda_0$ and $R_0$ are positive. This ensures that on 
classical solutions $M$ is positive. Also $N_1\ge0$. Here
$O(r^n)$ stands for a term whose magnitude at $r\to0$ is bounded by
$r^n$ times a constant, and whose $k$'th derivative at $r\to0$ is
similarly bounded by $r^{n-k}$ times a constant for $1\le k\le n$. 
The fact that the shift $N^r(t,r)$ vanishes as $r\to 0$ implies that
on solutions, the inner timelike boundary lies along a constant
Killing-time surface located completely in the past and the future
dynamical regions and cutting across the bifurcation two-sphere.
Also, on solutions, the variable $R_0$ corresponds to the throat-radius.

It is straightforward to verify that the conditions (\ref{s-r}) are
consistent with the equations of motion: Provided that the 
constraints obey $H=0=H^r$ and the fall-off conditions
(\ref{s-r-Lambda})--(\ref{s-r-PR}) hold for the initial data, and
provided that the lapse and shift satisfy (\ref{s-r-N})
and~(\ref{s-r-Nr}), it then follows that the fall-off conditions
(\ref{s-r-Lambda})--(\ref{s-r-PR}) are preserved in time by the 
time-evolution equations.

On the other hand, at $r=1$, the boundary conditions are as follows:
We fix $R$ and 
$-g_{tt} = N^2 - {(\Lambda N^r)}^2$ to be
prescribed positive-valued functions of~$t$. This means fixing the
metric on the three-surface $r=1$ to be timelike. In the classical 
solutions, the surface $r=1$ is located in the right exterior region 
of the Kruskal extension of the Schwarzschild spacetime.

We now give an action principle appropriate for these
boundary conditions. To begin, note that the surface action
$S_\Sigma [\Lambda, R, P_\Lambda, P_R ; N, N^r]$ in (\ref{S-ham}) is
well defined under the above conditions.
Consider the total action
\begin{equation}
S [\Lambda, R, P_\Lambda, P_R ; N, N^r]
= S_\Sigma [\Lambda, R, P_\Lambda, P_R ; N, N^r]
+ S_{\partial\Sigma} [\Lambda, R, P_\Lambda, P_R ; N, N^r]
\ \ ,
\label{S-total}
\end{equation}
where the boundary action is given by
\begin{eqnarray}
&&S_{\partial\Sigma} [\Lambda, R, P_\Lambda, P_R ; N, N^r]
\nonumber
\\
&&=
\int dt {\Biggl[ N R R' \Lambda^{-1} - N^r \Lambda P_\Lambda
- \case{1}{2} R {\dot R}
\ln \left|
{ N + \Lambda N^r \over  N - \Lambda N^r} \right|
\Biggr]}_{r=1}
\ \ ,
\label{S-boundary}
\end{eqnarray}
where $\left[{\it term}\right]_a$ is value of the {\it term} 
evaluated at $r=a$.
The variation of the total action (\ref{S-total}) can be written as a
sum of a volume term proportional to the equations of motion,
boundary terms from the initial and final spatial surfaces, and
boundary terms from $r=0$ and $r=1$. The boundary terms from the
initial and final spatial surfaces take the usual form
\begin{equation}
\pm \int_0^1 dr \, ( P_\Lambda \delta \Lambda + P_R \delta R )
\ \ ,
\label{bt-if}
\end{equation}
with the upper (lower) sign corresponding to the final (initial)
surface. These terms vanish provided we fix the initial and final
three-metrics. The boundary term from $r=0$ vanishes under the 
fall-off conditions specified in (\ref{s-r}). As will be shown in 
subsection \ref{subsec:reduction}, this is crucial in obtaining a reduced
Hamiltonian that corresponds to the correct quasilocal energy. The 
boundary term from $r=1$ reads
\begin{eqnarray}
\int dt \Bigg[ && \left( -P_R N^r + \Lambda^{-1} {(NR)}' \right)
\delta R
- \case{1}{2}
\ln \left|
{ N + \Lambda N^r \over  N - \Lambda N^r} \right|
\, \delta (R {\dot R})
\nonumber
\\
&&+ \case{1}{2} N^{-1} R
\left( \Lambda N^r {\dot R}
{\left( N^2 - {(\Lambda N^r)}^2 \right)}^{-1}
+ \Lambda^{-1} R' \right)
\delta \! \left( N^2 - {(\Lambda N^r)}^2 \right)
\nonumber
\\
&&{- \left( P_\Lambda + N^{-1}R ( {\dot R} - R'N^r ) \right)
\delta ( \Lambda N^r )
\Bigg]}_{r=1}
\ \ .
\label{bt-1}
\end{eqnarray}
Since $R$ and $N^2 - {(\Lambda N^r)}^2$ are fixed at $r=1$, the 
first three terms in (\ref{bt-1}) vanish. The integrand in the last 
term in (\ref{bt-1}) is proportional to the equation of
motion~(\ref{PLambda}), which is classically enforced for $0<r<1$ by
the volume term in the variation of the action. Therefore, for classical 
solutions, also the last term in (\ref{bt-1}) will vanish by continuity.

Thus the action (\ref{S-total}) is appropriate
for a variational principle which fixes the initial and final
three-metrics, and the three-metric on the timelike boundary at $r=1$.
Each classical solution belongs to that region of
a Kruskal diagram that lies within two timelike boundaries such that 
the inner boundary lies along a constant Killing-time surface located
completely in the dynamical regions and the outer boundary is a
timelike surface located in the right exterior region (see Fig. 3).
The constant $t$ slices are spacelike everywhere between the 
two timelike boundaries.

\subsubsection{Canonical transformation}
\label{subsec:transformation}

To obtain the reduced action and extract the unconstrained Hamiltonian 
system one needs to first solve the constraints (\ref{superham}) and 
(\ref{supermom}). In the following, we will follow Kucha\v{r}'s way of
handling the constraints \cite{KVK}. It was shown by Kucha\v{r} that in 
the context of a vacuum Schwarzschild spacetime (in the absence of 
timelike boundaries)
there exists a set of new variables, which are related to the ADM
variables through a canonical transformation, such that in terms of 
the new variables the constraints are remarkably simple and solvable.
This allows one to perform a Hamiltonian reduction.
In this section we show that the canonical transformation given by Kucha\v{r}
from the ADM variables $\left\{\Lambda , P_\Lambda ; R, P_R\right\}$
to the new variables $\left\{M, P_M ; {\sf R} , P_{\sf R}\right\}$ is
readily adapted to our boundary conditions. As mentioned earlier,
the boundary conditions ensure that $M>0$.

Recall from \cite{KVK} that the new variables
$\left\{M, P_M ; {\sf R} , P_{\sf R}\right\}$
are defined by
\begin{mathletters}
\label{trans}
\begin{eqnarray}
M &&= \case{1}{2} R (1-F)
\ \ ,
\label{trans-M}
\\
P_M &&= R^{-1} F^{-1} \Lambda P_\Lambda
\ \ ,
\label{trans-PM}
\\
{\sf R} &&= R
\ \ ,
\label{trans-sfR}
\\
P_{\sf R} &&=
P_R
- \case{1}{2} R^{-1} \Lambda P_\Lambda
- \case{1}{2} R^{-1} F^{-1} \Lambda P_\Lambda
\nonumber
\\
&& \; - R^{-1} \Lambda^{-2} F^{-1}
\left(
{(\Lambda P_\Lambda)}' (RR')
- (\Lambda P_\Lambda) {(RR')}'
\right)
\ \ ,
\label{trans-PsfR}
\end{eqnarray}
\end{mathletters}%
where
\begin{equation}
F = {\left( {R' \over \Lambda} \right)}^2
- {\left( { P_\Lambda \over R} \right)}^2
\ \ .
\label{F-def}
\end{equation}
In the classical solution, $M$ is the value of the Schwarzschild mass
and $-P_M$ is the derivative of the Killing time with respect to~$r$.
A pair of quantities which will become new Lagrange multipliers are
defined by
\begin{mathletters}
\label{N-def}
\begin{eqnarray}
{\sf N} &=&
{(4M)}^{-1}
\left(
N F^{-1} \Lambda^{-1} R'
- N^r R^{-1} F^{-1} \Lambda P_\Lambda
\right)
\ \ ,
\label{sfN-def}
\\
N^{\sf R} &=&
N^r R'
- N R^{-1} P_\Lambda
\ \ .
\label{NsfR-def}
\end{eqnarray}
\end{mathletters}%
Using arguments similar to Kucha\v{r} and LW, it can be shown that 
under our boundary conditions the transformation (\ref{trans}) is a 
canonical transformation, which is also invertible \cite{KVK}.

The Hamiltonian action (\ref{S-ham}) can now be written in terms of the 
new variables. Using Eqs.~(\ref{N-def}), one sees that the constraint terms
$NH + N^r H_r$ in the old surface action (\ref{S-ham}) take the form
$-4{\sf N} MM' + N^{\sf R} P_{\sf R}$. Thus the new surface action is
\begin{equation}
S_\Sigma
[M, {\sf R}, P_M, P_{\sf R} ; {\sf N} , N^{\sf R}]
= \int dt
\int _{0}^{1} dr
\left( P_M { \dot M}  +
P_{\sf R} \dot{\sf R} + 4 {\sf N} M M'
-N^{\sf R} P_{\sf R} \right)
\ \ ,
\label{S2-ham}
\end{equation}
where the quantities to be varied independently are $M$, ${\sf R}$,
$P_M$, $P_{\sf R}$, ${\sf N}$, and~$N^{\sf R}$.
The complete set of equations of motion is
\begin{mathletters}
\label{eom2}
\begin{eqnarray}
{\dot M} &=& 0
\ \ ,
\\
{\dot {\sf R}} &=& N^{\sf R}
\ \ ,
\label{eom2-R}
\\
{\dot P}_M &=& - 4M {\sf N}'
\ \ ,
\\
{\dot P}_{\sf R} &=& 0
\ \ ,
\\
MM' &=& 0
\ \ ,
\label{eom2-MM'}
\\
P_{\sf R} &=& 0
\ \ .
\label{eom2-PsfR}
\end{eqnarray}
\end{mathletters}%

We now turn to the boundary conditions and boundary terms.
As a preparation for this, let us denote by $Q^2$ the quantity
$-g_{tt}$ when expressed as a function of
the new canonical variables
and Lagrange multipliers. A short calculation using
(\ref{trans})--(\ref{N-def}) yields
\begin{equation}
Q^2 = - g_{tt} = 16M^2{\sf F} {\sf N}^2 - {\sf F}^{-1}
{\left(N^{\sf R}\right)}^2
\ \ .
\label{Q2}
\end{equation}
In general, $Q^2$ need not be positive for all values of $r$,
even for classical solutions.
However, as in subsection~\ref{subsec:BYgeo},
we shall introduce boundary conditions that fix the intrinsic metric
on the three-surface $r=1$ to be timelike, and under such 
boundary conditions $Q^2$ is positive at $r=1$. 

Consider now the total action
\begin{equation}
S [M, {\sf R}, P_M, P_{\sf R} ; {\sf N} , N^{\sf R}]
= S_\Sigma
[M, {\sf R}, P_M, P_{\sf R} ; {\sf N} , N^{\sf R}]
+ S_{\partial\Sigma}
[M, {\sf R}, P_M, P_{\sf R} ; {\sf N} , N^{\sf R}]
\ \ ,
\label{S2-total}
\end{equation}
where the boundary action is given by
\begin{eqnarray}
S_{\partial\Sigma}
[M, {\sf R}, P_M, P_{\sf R} ; {\sf N} , N^{\sf R}]
\nonumber
\\
&=&
\int dt {\left[
{\sf R} \sqrt{{\sf F}Q^2 + {\dot{\sf R}}^2}
+ \case{1}{2} {\sf R} \dot{\sf R}
\ln \left(
{\sqrt{{\sf F}Q^2 + {\dot{\sf R}}^2} - \dot{\sf R}
\over
\sqrt{{\sf F}Q^2 + {\dot{\sf R}}^2} + \dot{\sf R}
}
\right)
\vphantom{
{\left|
{\sqrt{{\sf F}Q^2 + {\dot{\sf R}}^2} - \dot{\sf R}
\over
\sqrt{{\sf F}Q^2 + {\dot{\sf R}}^2} + \dot{\sf R}
}
\right|}^Q_Q
}
\right]}_{r=1}
\label{S2-boundary}
\end{eqnarray}
with ${\sf F}=1-2M {\sf R}^{-1}$.
Note that the argument
of the logarithm in (\ref{S2-boundary}) is always
positive. The variation of (\ref{S2-total}) contains
a volume term proportional to the equations of motion, as
well as several boundary terms.
{}From the initial and final spatial
surfaces one gets the usual boundary terms
\begin{equation}
\pm \int _{0}^{1} dr
\left( P_M \delta M
+ P_{\sf R} \delta {\sf R}
\right)
\ \ ,
\end{equation}
which vanish provided we fix $M$ and ${\sf R}$ on these surfaces.
Similarly one can show that with our choice of boundary conditions
(given in section \ref{subsec:BYgeo}) the remaining boundary terms from 
the timelike surfaces at $r=0$ and $r=1$ also vanish.

\subsubsection{Hamiltonian reduction: the Brown-York Hamiltonian}
\label{subsec:reduction}

We now concentrate on the variational principle associated with the
action $S [M, {\sf R}, P_M, P_{\sf R} ; {\sf N} , N^{\sf
R}]$~(\ref{S2-total}). We shall reduce the
action to the true dynamical degrees of freedom by solving the
constraints.

The constraint $MM'=0$ (\ref{eom2-MM'}) implies that $M$ is
independent of~$r$. We can therefore write
\begin{equation}
M(t,r) = {\bf m}(t)
\ \ .
\label{bfm}
\end{equation}
Substituting this and the constraint $P_{\sf R}=0$
(\ref{eom2-PsfR}) back into
(\ref{S2-total}) yields the true Hamiltonian action
\begin{equation}
S [ {\bf m}, {\bf p} ; {\sf N_0} ;
{\sf R}_{\rm B}, Q_{\rm B} ] =
\int dt
\left(
{\bf p} {\dot {\bf m}}
- {\bf H} \right)
\ \ ,
\label{S-red}
\end{equation}
where
\begin{equation}
{\bf p} = \int_0^1 dr \, P_M
\ \ .
\label{bfp}
\end{equation}
The reduced Hamiltonian ${\bf H}$ in (\ref{S-red}) takes the form
\begin{equation}
\label{bfhB}
{\bf H} = - {\sf R}_{\rm B}
\sqrt{{\sf F}_{\rm B}Q_{\rm B}^2
+ {\dot {\sf R}}_{\rm B}^2}
- \case{1}{2} {\sf R}_{\rm B} {\dot {\sf R}}_{\rm B}
\ln \left(
{\sqrt{{\sf F}_{\rm B}Q_{\rm B}^2
+ {\dot {\sf R}}_{\rm B}^2} - {\dot {\sf R}}_{\rm B}
\over
\sqrt{{\sf F}_{\rm B}Q_{\rm B}^2
+ {\dot {\sf R}}_{\rm B}^2} + {\dot {\sf R}}_{\rm B}
}
\right)
\ \ .
\end{equation}
Here ${\sf R}_{\rm B}$ and $Q_{\rm B}^2$ are the values of ${\sf R}$
and $Q^2$ at the timelike boundary $r=1$, 
and ${\sf F} = 1 -
2 {\bf m} {\sf R}^{-1}$. As mentioned before, ${\sf R}_{\rm B}$ 
and $Q_{\rm B}^2$ are prescribed functions of time,
satisfying ${\sf R}_{\rm B}>0$ and $Q_{\rm B}^2>0$.
Note that ${\bf H}$ is, in general, explicitly time-dependent.

The variational principle associated with the reduced
action~(\ref{S-red}) fixes the initial and final values of~${\bf m}$.
The equations of motion are
\begin{mathletters}
\label{red-eom}
\begin{eqnarray}
{\dot {\bf m}} &=& 0
\ \ ,
\label{red-eom1}
\\
\noalign{\smallskip}
{\dot {\bf p}} &=&
- {\partial {\bf H} \over \partial {\bf m}}
\nonumber
\\
&=&
- {\sf F}_{\rm B}^{-1}
\sqrt{{\sf F}_{\rm B}Q_{\rm B}^2
+ {\dot {\sf R}}_{\rm B}^2}
\ \ .
\label{red-eom2}
\end{eqnarray}
\end{mathletters}%
Equation (\ref{red-eom1}) is readily understood in terms of the
statement that ${\bf m}$ is classically equal to the time-independent
value of the Schwarzschild mass. To interpret
equation~(\ref{red-eom2}), recall from
Sec.~\ref{subsec:transformation} that $-P_M$ equals classically the
derivative of the Killing time with respect to~$r$, and ${\bf p}$
therefore equals by (\ref{bfp}) the difference of the Killing times
at the left and right ends of the constant $t$ surface. As the
constant $t$ surface evolves in the Schwarzschild spacetime, 
(\ref{red-eom2}) gives the negative of the evolution rate of the
Killing time at the right end of the spatial surface where it
terminates at the outer timelike boundary at $r=1$. Note that 
${\dot {\bf p}}$ gets no contribution from the inner timelike boundary 
located at $r=0$ in the dynamical region. This is a consequence of the 
fall-off conditions (\ref{s-r}) which ensure that on solutions, the rate 
of evolution of the Killing time at $r=0$ is zero.

The case of interest is when the radius of the `outer' boundary 
two-sphere does not change in time, i.e., 
${\dot {\sf R}}_{\rm B}=0$. In that case the second term in 
${\bf H}_{\rm B}$ (\ref{bfhB}) vanishes, and ${\dot {\bf p}}$ in 
(\ref{red-eom2})
is readily understood in terms of the Killing time of a static
Schwarzschild observer, expressed as a function of the proper time
$\int^t dt' \sqrt{Q_{\rm B}^2(t')}$ and the blueshift factor ${\sf
F}_{\rm B}^{-1/2}$. The reduced Hamiltonian is given by
\begin{equation}
{\bf H}=
- B\sqrt{{\sf F}_{\rm B}Q_{\rm B}^2},
\end{equation}
where $B$ is the time-independent value of ${\sf R}_{\rm B}$.
Unfortunately, the above Hamiltonian does not vanish as ${\bf m}$ goes 
to zero. The situation is remedied by adding the $K_0=BQ_{\rm B}$ 
term of Gibbons and Hawking \cite{GH1} to ${\bf H}$. Physically, this
added term arises from the extrinsic curvature of the `outer' boundary 
two-sphere when embedded in flat spacetime. With the added term, the
Hamiltonian becomes
\begin{equation}
\label{BYHQ}
{\bf H} =\left(1- \sqrt{1-2{\bf m}B^{-1}}\right)BQ_{\rm B}.
\end{equation}
This is the quasilocal energy of Brown and York when $Q_B=1$. The choice
of $Q_B$ determines the choice of time in the above Hamiltonian. Setting
$Q_B=\sqrt{-g_{tt}}=1$ geometrically means choosing a spacetime 
foliation in which the rate of evolution of the spatial hypersurface 
on the box is the same as that of the proper time. Then the new Hamiltonian
is 
\begin{equation}
\label{BYH}
{\sf H}= \left(1- \sqrt{1-2{\bf m}B^{-1}}\right)B
\ \ ,
\end{equation}
namely, the quasilocal energy (\ref{quasiH}) in Schwarzschild spacetime. 
In the next section, we discuss the geometric relevance of the LW 
Hamiltonian that, as we showed earlier,
yields the correct free energy of the system.

\subsection{The Louko-Whiting Hamiltonian}
\label{subsec:LWgeoham}

We now summarize the LW choice of the foliation of the Schwarzschild spacetime,
state the corresponding boundary conditions they imposed, and briefly 
mention how they obtain their reduced Hamiltonian. The main purpose of this 
section is to facilitate a comparison between the LW boundary conditions
and our choice of the boundary conditions (as discussed in the preceeding
subsections) that yield the BY Hamiltonian. 

\subsubsection{Louko-Whiting boundary conditions}
\label{subsec:LWgeo}

As shown in Fig. 2, LW considered a foliation 
in which the spatial hypersurfaces are restricted to lie in the right
exterior region of the Kruskal diagram. Each spatial hypersurface in this
region extends from the box at the right end up to the bifurcation
2-sphere on the left end.

The boundary conditions imposed by LW are as follows. At $r\to0$, they 
adopt the fall-off conditions
\begin{mathletters}
\label{ls-r}
\begin{eqnarray}
\Lambda (t,r) &=& \Lambda_0(t) + O(r^2)
\ \ ,
\label{ls-r-Lambda}
\\
R(t,r) &=& R_0(t) + R_2(t) r^2 + O(r^4)
\ \ ,
\label{ls-r-R}
\\
P_{\Lambda}(t,r) &=& O(r^3)
\ \ ,
\label{ls-r-PLambda}
\\
P_{R}(t,r) &=& O(r)
\ \ ,
\label{ls-r-PR}
\\
N(t,r) &=& N_1(t)r + O(r^3)
\ \ ,
\label{ls-r-N}
\\
N^r(t,r) &=& N^r_1(t)r + O(r^3)
\ \ ,
\label{ls-r-Nr}
\end{eqnarray}
\end{mathletters}%
where $\Lambda_0$ and $R_0$ are positive, and $N_1\ge0$.
Equations (\ref{ls-r-Lambda})
and (\ref{ls-r-R}) imply that the classical solutions have a positive
value of the Schwarzschild mass, and that the constant $t$ slices at
$r\to0$ are asymptotic to surfaces of constant Killing time in the
right hand side exterior region in the Kruskal diagram, all
approaching the bifurcation two-sphere as $r\to0$. The spacetime
metric has thus a coordinate singularity at $r\to0$, but this
singularity is quite precisely controlled. In particular, on a
classical solution the future unit normal to a constant $t$ surface
defines at $r\to0$ a future timelike unit vector $n^a(t)$ at the
bifurcation two-sphere of the Schwarzschild spacetime, and the evolution of
the constant $t$ surfaces boosts this vector at the rate given by
\begin{equation}
n^a(t_1) n_a(t_2) =
-\cosh\left(\int_{t_1}^{t_2}
\Lambda_0^{-1}(t) N_1(t) \, dt \right)
\ \ .
\label{n-boost}
\end{equation}

At $r=1$, we fix $R$ and $-g_{tt} = N^2 - {(\Lambda N^r)}^2$ to be
prescribed positive-valued functions of~$t$. This means fixing the
metric on the three-surface $r=1$, and in particular fixing this
metric to be timelike. In the classical solutions, the surface $r=1$
is located in the right hand side exterior region of the Kruskal diagram.

To obtain an action principle appropriate for these
boundary conditions, consider the total action
\begin{equation}
S [\Lambda, R, P_\Lambda, P_R ; N, N^r]
= S_\Sigma [\Lambda, R, P_\Lambda, P_R ; N, N^r]
+ S_{\partial\Sigma} [\Lambda, R, P_\Lambda, P_R ; N, N^r]
\ \ ,
\label{LS-total}
\end{equation}
where the boundary action is given by
\begin{eqnarray}
&&S_{\partial\Sigma} [\Lambda, R, P_\Lambda, P_R ; N, N^r]
\nonumber
\\
&&=
\case{1}{2} \int dt \, {\left[ R^2 N' \Lambda^{-1} \right]}_{r=0}
\; \;
+ \int dt {\Biggl[ N R R' \Lambda^{-1} - N^r \Lambda P_\Lambda
- \case{1}{2} R {\dot R}
\ln \left|
{ N + \Lambda N^r \over  N - \Lambda N^r} \right|
\Biggr]}_{r=1}
\ \ .
\label{LS-boundary}
\end{eqnarray}
The variation of the total action (\ref{LS-total}) can be written as a
sum of a volume term proportional to the equations of motion,
boundary terms from the initial and final spatial surfaces, and
boundary terms from $r=0$ and $r=1$. 

To make the action (\ref{LS-total}) appropriate
for a variational principle, one fixes the initial and final
three-metrics, the box-radius $R$, and the three-metric on the timelike 
boundary at $r=1$. These are similar to the boundary conditions that we 
imposed to obtain the BY Hamiltonian. However, for the LW boundary 
conditions, one has to also fix the quantity ${\bf N}_0 \equiv 
N_1 \Lambda_0^{-1}= \lim_{r\to0} N' \Lambda^{-1}$ at the
bifurcation 2-sphere.
Each classical solution is part of the right hand exterior region of
a Kruskal diagram, with the constant $t$ slices approaching the
bifurcation two-sphere as $r\to0$, and $N_1 \Lambda_0^{-1}$ giving
via (\ref{n-boost}) the rate of change of the unit normal to the
constant $t$ surfaces at the bifurcation two-sphere.

Although we
are here using geometrized units, the
argument of the $\cosh$ in (\ref{n-boost})
is a truly dimensionless
``boost parameter" even in physical units.

\subsubsection{Hamiltonian reduction: the Louko-Whiting Hamiltonian}
\label{subsec:LWreduction}

To obtain the (reduced) LW Hamiltonian, one needs to solve the 
super-Hamiltonian and the supermomentum constraints. Just as in the
case of the BY Hamiltonian (see subsection \ref{subsec:transformation}),
it helps to first make a canonical transformation to the Kucha\v{r} 
variables (see LW for details). After solving the constraints, one obtains 
the  following reduced action
\begin{equation}
S [ {\bf m}, {\bf p} ; {\bf N_0} ;
{\sf R}_{\rm B}, Q_{\rm B} ] =
\int dt
\left(
{\bf p} {\dot {\bf m}}
- {\bf h} \right)
\ \ ,
\label{LS-red}
\end{equation}
where
\begin{eqnarray}
{\bf p} &=& \int_0^1 dr \> P_M
\ \ ,
\\
{\bf m}(t) = M(t,r)
\ \ ,
\label{Lbfp}
\end{eqnarray}
and the reduced Hamiltonian ${\bf h}$ in (\ref{LS-red}) is
\begin{equation}
{\bf h}(t) = \left( 1 -\sqrt{1-{2 {\bf m} \over B}}\right) B Q_B(t) - 
          2{\bf N}_0 (t) {\bf m}^2
\ \ , 
\end{equation}
which is the same as the one given in Eq. (\ref{LWht}). In obtaining the
above reduced form ${\bf h(t)}$, we have assumed that the box-radius is 
constant in time, $\dot{R}_B =0$, just as we 
did in obtaining the BY Hamiltonian (\ref{BYHQ}).

\section{geometric origins of thermodynamic parameters}
\label{geothermo}

Having established the geometrical significance of the BY and LW 
Hamiltonians, the basis for their thermodynamical roles becomes apparent.
We showed that the BY Hamiltonian evolves spatial hypersurfaces in such a way 
that they span the spacetime region both inside and outside the event 
horizon. This is what one would expect from the fact that it corresponds 
to the quasilocal energy of the complete spacetime region enclosed inside
the box. On the other hand, the LW Hamiltonian evolves spatial slices that
are restricted to lie outside the event horizon. With our choice of the 
boost parameter ${\bf N}_0$, this corresponds to the Helmholtz free energy
of the system, which is less than the quasilocal energy: this is expected 
since the LW slices span a smaller region of the spacetime compared to 
the BY slices. Also, the fact that the LW slices are limited to 
lie outside the event horizon implies that the energy on these slices can
be harnessed by an observer located outside the box. This is consistent
with the fact that it corresponds to the Helmholtz free energy of the 
system -- which is the amount of energy in a system that is available 
for doing work by the system on its surroundings.

Using the thermodynamical roles played by the LW Hamiltonian and the BY 
Hamiltonian (see section \ref{sec:thermoconsi}), we now  
derive, at the classical level, many of the thermodynamical quantities 
associated with the Schwarzschild black hole enclosed inside a box. We
begin by finding the temperature on the box. From Eq. (\ref{LWh}), the
Helmholtz free energy is
\begin{equation}
\label{LWhF1}
{\cal F}={\sf h} = \left( 1 -\sqrt{1-{2{\bf m}\over B}}\right) B  - 
          2{\bf N}_0  {\bf m}^2, 
\end{equation}
The above equation, along with Eqs. (\ref {F}) and (\ref{E}),
implies that 
\begin{equation}
\label{TS}
{\cal T}{\cal S}\> = \> 2 {\bf N}_0 {\bf m}^2,
\end{equation}
or,
\begin{equation}
\label{S1}
{\cal S}\> = \> 2 {\bf N}_0 {\bf m}^2 \beta,
\end{equation}
where $\beta\equiv {\cal T}^{-1}$. Equation (\ref{S1}) gives an expression 
for the entropy in terms of the geometrical quantity ${\bf N}_0$. 
On the other hand one can find ${\cal S}$ 
also from the thermodynamic identity
\begin{equation}
\label{S2}
{\cal S}\> =\> \left( 1 - \beta {\partial \over\partial\beta}\right) \ln Z,
\end{equation}
where $Z$ is the partition function defined by Eq. (\ref{partitionh})
and Eq. (\ref{LWhF1}). Then Eq. (\ref{S2}) gives the entropy to be
\begin{equation}
\label{S3}
{\cal S}\> = -\> 2 {\bf m}^2 \beta^2 {\partial {\bf N}_0\over\partial\beta}.
\end{equation}
The above equation gives another expression for the entropy, now in terms of 
the derivative of ${\bf N}_0$. Comparing Eqs. (\ref{S1}) and (\ref{S3}) 
we find
\begin{equation}
\label{N0beta}
{\bf N}_0 = \> q \beta^{-1} = q {\cal T},
\end{equation}
where $q$ is some undetermined quantity that is independent of $\beta$. 

The exact form of ${\cal T}$ as a function of ${\bf m}$ is
found by noting that the free energy ${\cal F}$ should be a minimum at 
equilibrium. Since in a canonical ensemble the box-radius $R$ and the 
temperature ${\cal T}$ (which is proportional to ${\bf N}_0$) are fixed, 
the only quantity in ${\cal F}$ that can vary is ${\bf m}$. 
Thus the  question we ask is the following: For a
fixed value of the curvature radius $R=B$ and the boost parameter 
${\bf N}_0$, what is the value of ${\bf m}$ that minimizes  ${\cal F}$?
>From the expression for ${\cal F}$ in Eq. (\ref{LWhF1})
one finds this value of ${\bf m}$, to be a function of ${\bf N}_0$. 
Inverting this relation gives
\begin{equation}
\label{N0m}
{\bf N}_0 = \left(4 {\bf m} \sqrt{1-{2{\bf m} \over B}}\right)^{-1}.
\end{equation}
>From (\ref{N0beta}) and (\ref{N0m}) we find that the equilibrium temperature 
on a box of radius $B$ enclosing a black hole of mass ${\bf m}$ obeys 
\begin{equation}
\label{beta}
{\cal T} \propto \left(4 {\bf m}\sqrt{1-{2 {\bf m} \over B}}\right)^{-1}
\ \ ,
\end{equation}
in agreement with known results. Significantly, Eq. (\ref{N0beta}) shows
that the equilibrium temperature geometrically corresponds to a particular
value of the boost parameter.

>From Eqs. (\ref{S1}) and (\ref{N0beta}) we find that the entropy of a 
Schwarzschild black hole is quadratic in its mass. 
Unfortunately, in this formalism one can not determine the correct
constants of proportionality in ${\cal S}$ and ${\cal T}$. However, notice
that our derivation is purely classical. Although simple mathematically, this
derivation is incomplete due to the lack of the constant of proportionality 
$q$ in Eq. (\ref{N0beta}). The correct value for this constant, 
$q=2\pi/ \hbar$, can be obtained only from a quantum treatment.

Finally, we note that for the spatial slices that obey 
${\bf N}_0 = (4{\bf m} \sqrt{1-2 {\bf m} / B })^{-1}$, the free energy 
can be obtained from (\ref{LWh}) to be
\begin{equation}
{\cal F} = \left( 1 - \sqrt{1- {2{\bf m} \over B}}\right) B - {{\bf m} \over 2
      \sqrt{1-{2{\bf m} \over B}}}.
\end{equation}
The above equation shows that if the radius of the box $B$ is 
kept fixed, then the free energy of the system is minimum for 
the configuration with a black hole of mass ${\bf m} = B/3$.

\section{Conclusions}
\label{sec:conclu}

In this work our goal was to seek a geometrical basis for the 
thermodynamical
aspects of a black hole. We find that the value of the Brown-York Hamiltonian 
can be interpreted as the internal energy of a black hole inside a box.
Whereas the value of the Louko-Whiting Hamiltonian gives the Helmholtz 
free energy of the system. After finding these thermodynamical roles 
played by the BY and LW Hamiltonians, we ask what the geometrical 
significance of these Hamiltonians is. 

In this regard the geometrical role of the LW Hamiltonian was already known.
It was recently shown by LW that their Hamiltonian evolves spatial 
hypersurfaces in a special foliation of the Kruskal diagram. The 
characteristic feature of this foliation is that it is
limited to only the right exterior region of this spacetime (see Fig. 2)
and the spatial hypersurfaces are required to converge onto the bifurcation
2-sphere, which acts as their inner boundary (the box itself being the
outer boundary). 

On the other hand, the geometrical significance of the BY Hamiltonian 
as applied to the black hole case was not fully known, although it had 
been argued that its value is the energy of the Schwarzschild spacetime
region that is enclosed inside a spherical box. In this work we establish
the geometric role of the BY Hamiltonian by showing that it is the generator
of time-evolution of spatial hypersurfaces in certain foliations of the
Schwarzschild spacetime.

Establishing the thermodynamic connection of the BY and LW Hamiltonians 
allowed us to obtain a geometrical interpretation for the 
equilibrium temperature
of a black hole enclosed inside a box, i.e., as measured by a stationary 
observer on the box. Geometrically, the temperature turns out to be the 
rate at which the LW spatial hypersurfaces are boosted at the bifurcation
2-sphere. One could however ask what happens if the LW hypersurfaces are
evolved at a different rate, i.e., if the label time  $t$ is chosen to be
boosted with respect to the proper time of a stationary observer on the box.
In that case, it can be shown that the BY Hamiltonian and the rate at 
which the LW hypersurfaces are evolved at the bifurcation 2-sphere get 
``blue-shifted'' by the appropriate boost-factor. On the other hand,
the entropy of the system can still be interpreted as the change in free
energy per unit change in the temperature of the system.

\section{Acknowledgments}

We thank Abhay Ashtekar, Viqar Husain, Eric Martinez, Jorg\'{e} Pullin, 
Lee Smolin, and Jim York for helpful discussions. We would especially like 
to thank Jorma Louko for critically reading the manuscript and making 
valuable comments. Financial support from IUCAA is gratefully acknowledged
by one of us (SB). This work was supported in part by NSF Grant No. 
PHY-95-07740.

\appendix\section{The Witten black hole}
\label{sec:2d}

The approach we describe above in studying the thermodynamics of 
4D spherically symmetric Einstein gravity can also be extended to the
case of the 2D vacuum dilatonic black hole \cite{Witten} in an analogous 
fashion. In the case of a 2D black hole, the event horizon is located
at a curvature radius $R_{EH} = {\bf m}/(2\lambda^2)$, 
where $\lambda^{-1}$ is
a positive constant that sets the length-scale in the 2D models.
The quasilocal energy of a system comprising 
of such a black hole in the presence of a timelike boundary situated at 
a curvature radius $B$ can be shown to be 
\begin{equation}
\label{app:quasi2H}
{}^2 {\sf H} =  \left( 1 - \sqrt{1-{{\bf m} \over 2\lambda^2 B}}\right) 
                                       4\lambda^2 B, 
\end{equation}
which strongly resembles the 4D counterpart in (\ref{quasiH}). ${}^2 {\sf H}$
evolves constant $t$ spatial hypersurfaces that extend from an inner
timelike boundary lying on a constant Killing-time surface in the dynamical
region up to a timelike boundary (the box) placed in the right exterior 
region (see Fig. 3). 
 
  The Hamiltonian that evolves the two-dimensional counterpart of the 
Louko and Whiting spatial slices that extend from the bifurcation point
up to the box (see Fig. 2) is 
\begin{equation}
\label{app:BLPPht}
{}^2 {\bf h}(t) = \left( 1 -\sqrt{1-{{\bf m} \over 2\lambda^2 B}}\right)
                 4\lambda^2 B Q_B(t)
        - \> {}^2 {\bf N}_0 (t) (2\pi){\bf m} \lambda^{-1}, 
\end{equation}
where in general $Q_B$ and ${}^2 {\bf N}_0$ are functions of time. The above
Hamiltonian ${}^2 {\bf h}$ was found in Ref. 
\cite{BLPP}. 
There it was found that ${\bf N}^M_0\equiv - {}^2 {\bf N}_0(2\pi) 
\lambda^{-1}$
is the rate at which the spatial hypersurface are boosted at the 
bifurcation point. On the other hand, $Q_B\equiv \sqrt{-g_{tt}}$, 
$(-g_{tt})$ being the
time-time component of the spacetime metric on the box.  
If one restricts the spatial
hypersurfaces to approach the box along constant proper-time hypersurfaces, 
then, as in 4D, $Q_B=1$. 
Using fall-off conditions on the ADM variables at the bifurcation point 
analogous to the LW fall-off conditions (\ref{ls-r}), it can be shown that 
on solutions \cite{BLPP} ${}^2 {\bf N}_0 = \kappa dT/dt$ where 
$T$ is the Killing time, and $\kappa = \lambda/(2\pi)$ is the surface 
gravity of a Witten black hole.
The time-evolution of these restricted spatial hypersurfaces is given
by the Hamiltonian 
\begin{equation}
\label{app:BLPPh}
{}^2{\sf h} =  \left( 1 -\sqrt{1-{{\bf m}\over 2\lambda^2 B}}\right)
                 4\lambda^2 B  
          -{}^2 {\bf N}_0  {\bf m}(2\pi)\lambda^{-1} . 
\end{equation}

Like the 4D case, here too it can be shown that
${}^2 {\sf H}$ is analogous to the internal energy, whereas ${}^2 {\sf h}$
denotes the Helmholtz free energy of the 2D system. A similar analysis
also shows that 
\begin{equation}
\label{app:2N0beta}
{}^2 {\bf N}_0 \propto \beta^{-1}
\end{equation}
and
\begin{equation}
\beta  = \left( {2\pi \over \lambda}
\sqrt{1-{{\bf m} \over 2\lambda^2 B}}
           \right)
\ \ ,
\end{equation}
which is inverse of the blue-shifted temperature on the box. The temperature 
of a 2D black hole at infinity on the other hand is $\lambda / (2\pi)$,
which is independent of the black hole mass.

\section{Brown-York Hamiltonian and time-evolution in 
a new foliation}
\label{app:quasiH2}

In Section \ref{sec:thermoconsi}, we found a choice of spatial hypersurfaces
that were evolved by the BY Hamiltonian under a specific set of
boundary conditions. In this appendix we find a different choice 
of spatial hypersurfaces, i.e., with a different inner boundary,
that is evolved by the BY Hamiltonian under a different set of 
boundary conditions. 

We begin by stating the boundary conditions and specifying the
spacetime foliation they define. At the inner boundary $r=0$, we 
fix $R$ and $g_{tt} = -N^2 + {(\Lambda N^r)}^2$ to be
prescribed positive-valued functions of~$t$. This means fixing the
metric on the three-surface $r=0$, and in particular fixing this
metric to be spacelike there. 
On the other hand, at $r=1$, we fix $R$ and 
$-g_{tt} = N^2 - {(\Lambda N^r)}^2$ to be
prescribed positive-valued functions of~$t$. This means fixing the
metric on the three-surface $r=1$ to be timelike. In the classical 
solutions, the surface $r=1$ is located in the right exterior region 
of the Kruskal diagram.

We now wish to give an action principle appropriate for these
boundary conditions. Note that the surface action
$S_\Sigma [\Lambda, R, P_\Lambda, P_R ; N, N^r]$ given by Eq. (\ref{S-ham}) 
is well defined under the above conditions. Consider the total action
\begin{equation}
S [\Lambda, R, P_\Lambda, P_R ; N, N^r]
= S_\Sigma [\Lambda, R, P_\Lambda, P_R ; N, N^r]
+ S_{\partial\Sigma} [\Lambda, R, P_\Lambda, P_R ; N, N^r]
\ \ ,
\label{app:S-total}
\end{equation}
where the boundary action is given by
\begin{eqnarray}
&&S_{\partial\Sigma} [\Lambda, R, P_\Lambda, P_R ; N, N^r]
\nonumber
\\
&&=
\int dt {\Biggl[ N R R' \Lambda^{-1} - N^r \Lambda P_\Lambda
- \case{1}{2} R {\dot R}
\ln \left|
{ N + \Lambda N^r \over  N - \Lambda N^r} \right|
\Biggr]}_{r=0}^{r=1}
\ \ ,
\label{app:S-boundary}
\end{eqnarray}
where $\left[{\it term}\right]_a^b$ implies the difference in the 
values of the {\it term} evaluated at $r=b$ and at $r=a$.
The variation of the total action (\ref{app:S-total}) can be written as a
sum of a volume term proportional to the equations of motion,
boundary terms from the initial and final spatial surfaces, and
boundary terms from $r=0$ and $r=1$. The boundary terms from the
initial and final spatial surfaces take the usual form
\begin{equation}
\pm \int_0^1 dr \, ( P_\Lambda \delta \Lambda + P_R \delta R )
\ \ ,
\label{app:bt-if}
\end{equation}
with the upper (lower) sign corresponding to the final (initial)
surface. These terms vanish provided we fix the initial and final
three-metrics. The boundary term from $r=0$ and $r=1$ read
\begin{eqnarray}
\int dt \Bigg[ && \left( -P_R N^r + \Lambda^{-1} {(NR)}' \right)
\delta R
- \case{1}{2}
\ln \left|
{ N + \Lambda N^r \over  N - \Lambda N^r} \right|
\, \delta (R {\dot R})
\nonumber
\\
&&+ \case{1}{2} N^{-1} R
\left( \Lambda N^r {\dot R}
{\left( N^2 - {(\Lambda N^r)}^2 \right)}^{-1}
+ \Lambda^{-1} R' \right)
\delta \! \left( N^2 - {(\Lambda N^r)}^2 \right)
\nonumber
\\
&&{- \left( P_\Lambda + N^{-1}R ( {\dot R} - R'N^r ) \right)
\delta ( \Lambda N^r )
\Bigg]}_{r=0}^{r=1}
\ \ ,
\label{app:bt-1}
\end{eqnarray}
where $\left[{\it term}\right]^b_a$ implies the difference between the
values of the {\it term} evaluated at $r=b$ and at $r=a$.
As $R$ and $N^2 - {(\Lambda N^r)}^2$ are fixed at $r=0$ and $r=1$, the 
first three terms in (\ref{app:bt-1}) vanish. The integrand in the last 
term in (\ref{app:bt-1}) is proportional to the equation of
motion~(\ref{PLambda}), which is classically enforced for $0<r<1$ by
the volume term in the variation of the action.
Therefore, for classical solutions,
also the last term in (\ref{app:bt-1})
will vanish by continuity.

We thus conclude that the action (\ref{app:S-total}) is appropriate
for a variational principle which fixes the initial and final
three-metrics, the three-metric on the spacelike boundary at $r=0$
and the three-metric on the timelike boundary at $r=1$.
Each classical solution belongs to that region of
a Kruskal diagram that lies to the future of the null line at Killing
time $T\rightarrow -\infty$.
The constant $t$ slices are spacelike everywhere between the spacelike
boundary at $r=0$, and the timelike boundary at $r=1$.

\subsection{Canonical transformation}
\label{subsec:Atransformation}
		
The canonical transformation given in Kucha\v{r} \cite{KVK} from the 
variables $\left\{\Lambda , P_\Lambda ; R, P_R\right\}$
to the new variables $\left\{M, P_M ; {\sf R} , P_{\sf R}\right\}$ is
readily adapted to our boundary conditions. As mentioned earlier,
we shall assume that $M>0$. Recall that the new variables 
$\left\{M, P_M ; {\sf R} , P_{\sf R}\right\}$ 
have been defined in subsection \ref{subsec:transformation} 
by equations (\ref{trans}) and (\ref{F-def}). The new Lagrange multipliers 
are defined in Eq. (\ref{N-def}). 
It can be shown that the transformation (\ref{trans}) is a 
canonical transformation also under the new boundary conditions being 
considered in this appendix. 

We wish to write an action in terms of the new variables. Using
Eqs.~(\ref{N-def}), one finds that the constraint terms
$NH + N^r H_r$ in the old surface action (\ref{S-ham}) take the form
$-4{\sf N} MM' + N^{\sf R} P_{\sf R}$ and 
the new surface action is the same as that given in (\ref{S2-ham}).
Therefore, the equations of motion remain unchanged and are given
by (\ref{eom2}).

We now turn to the boundary conditions and boundary terms.
As before, we define
\begin{equation}
Q^2 = - g_{tt} = 16M^2{\sf F} {\sf N}^2 - {\sf F}^{-1}
{\left(N^{\sf R}\right)}^2
\ \ .
\label{app:Q2}
\end{equation}
In general, $Q^2$ need not be positive for all values of $r$,
even for classical solutions. As in the preceeding section,
we shall introduce boundary conditions that fix the intrinsic metric
of the three-surfaces $r=0$ and $r=1$ to be spacelike and timelike,
respectively, and under such 
boundary conditions $Q^2$ is negative at $r=0$ but positive at $r=1$. 
{}From (\ref{app:Q2}) it is then seen that ${\sf N}$ is nonzero at 
$r=0$ and $r=1$.
Recalling that we are assuming $N>0$, Eq.~(\ref{sfN-def})
shows that ${\sf N}$ is positive at $r=1$ for
classical solutions with the Schwarzschild slicing, since in this
slicing one has $P_\Lambda=0$. Continuity then implies that ${\sf N}$
must be positive at $r=1$ for all classical solutions compatible with
our boundary conditions. On the other hand, at $r=0$ we now put the 
additional condition that ${\sf F} < 0$ (or, equivalently, $F < 0$).
On classical solutions, this extra condition restricts the surface $r=0$
to lie either in the past or the future dynamical region of 
the Schwarzschild spacetime. Although the final expression for the 
quasilocal energy is independent of this choice, for definiteness
we will choose the spatial boundary at $r=0$ to lie in the future
dynamical region (see Fig. 4). Then at $r=0$, Eq.~(\ref{sfN-def}) shows that 
${\sf N}$ has to be negative because $F<0$ there. We can therefore, without 
loss of generality, choose to work in a neighborhood of the classical
solutions such that ${\sf N}$ is positive at $r=1$ whereas
${\sf N}$ is negative at $r=0$. 

Consider now the total action
\begin{equation}
S [M, {\sf R}, P_M, P_{\sf R} ; {\sf N} , N^{\sf R}]
= S_\Sigma
[M, {\sf R}, P_M, P_{\sf R} ; {\sf N} , N^{\sf R}]
+ S_{\partial\Sigma}
[M, {\sf R}, P_M, P_{\sf R} ; {\sf N} , N^{\sf R}]
\ \ ,
\label{app:S2-total}
\end{equation}
where the boundary action is given by
\begin{eqnarray}
S_{\partial\Sigma}
[M, {\sf R}, P_M, P_{\sf R} ; {\sf N} , N^{\sf R}]
\nonumber
\\
&=&
\int dt {\left[
{\sf R} \sqrt{{\sf F}Q^2 + {\dot{\sf R}}^2}
+ \case{1}{2} {\sf R} \dot{\sf R}
\ln \left(
{\sqrt{{\sf F}Q^2 + {\dot{\sf R}}^2} - \dot{\sf R}
\over
\sqrt{{\sf F}Q^2 + {\dot{\sf R}}^2} + \dot{\sf R}
}
\right)
\vphantom{
{\left|
{\sqrt{{\sf F}Q^2 + {\dot{\sf R}}^2} - \dot{\sf R}
\over
\sqrt{{\sf F}Q^2 + {\dot{\sf R}}^2} + \dot{\sf R}
}
\right|}^Q_Q
}
\right]}_{r=1}
\nonumber
\\
&-&\int dt {\left[
{\sf R} \sqrt{{\sf F}Q^2 + {\dot{\sf R}}^2}
+ \case{1}{2} {\sf R} \dot{\sf R}
\ln \left(
{\sqrt{{\sf F}Q^2 + {\dot{\sf R}}^2} - \dot{\sf R}
\over
\sqrt{{\sf F}Q^2 + {\dot{\sf R}}^2} + \dot{\sf R}
}
\right)
\vphantom{
{\left|
{\sqrt{{\sf F}Q^2 + {\dot{\sf R}}^2} - \dot{\sf R}
\over
\sqrt{{\sf F}Q^2 + {\dot{\sf R}}^2} + \dot{\sf R}
}
\right|}^Q_Q
}
\right]}_{r=0}
\label{app:S2-boundary}
\end{eqnarray}
with ${\sf F}=1-2M {\sf R}^{-1}$.
Note that the argument
of the logarithm in (\ref{app:S2-boundary}) is always
positive. The variation of (\ref{app:S2-total}) contains
a volume term proportional to the equations of motion, as
well as several boundary terms. These boundary terms vanish if
on the initial and final three-surfaces we fix the new canonical 
coordinates $M$ and~${\sf R}$, and at $r=0$ and $r=1$ we fix ${\sf R}$
and the intrinsic metric on these three-surfaces.

\subsection{Hamiltonian reduction}
\label{subsec:Areduction}

We now reduce the action $S [M, {\sf R}, P_M, P_{\sf R} ; {\sf N} , N^{\sf
R}]$~(\ref{S2-total}) to the true dynamical degrees of freedom by solving 
the constraints (\ref{eom2-MM'}) and (\ref{eom2-PsfR}) as before. This gives 
the true Hamiltonian action to be
\begin{equation}
S [ {\bf m}, {\bf p} ; {\sf N_0} ;
{\sf R}_{\rm B}, Q_{\rm B} ] =
\int dt
\left(
{\bf p} {\dot {\bf m}}
- {\bf H} \right)
\ \ ,
\label{app:S-red}
\end{equation}
where ${\bf p}$ and ${\bf m}$ are defined as in section \ref{subsec:reduction}.
The reduced Hamiltonian ${\bf h}$ in (\ref{app:S-red}) takes the form
\begin{equation}
{\bf H} = {\bf H}_{\rm S} + {\bf H}_{\rm B}
\ \ ,
\end{equation}
with
\begin{mathletters}
\begin{eqnarray}
{\bf H}_{\rm S} &=& 
- {\sf R}_{\rm S}
\sqrt{{\sf F}_{\rm S}Q_{\rm S}^2
+ {\dot {\sf R}}_{\rm S}^2}
- \case{1}{2} {\sf R}_{\rm S} {\dot {\sf R}}_{\rm S}
\ln \left(
{\sqrt{{\sf F}_{\rm S}Q_{\rm S}^2
+ {\dot {\sf R}}_{\rm S}^2} - {\dot {\sf R}}_{\rm S}
\over
\sqrt{{\sf F}_{\rm S}Q_{\rm S}^2
+ {\dot {\sf R}}_{\rm S}^2} + {\dot {\sf R}}_{\rm S}
}
\right)
\ \ ,
\label{app:bfhS}
\\
{\bf H}_{\rm B} &=&
{\sf R}_{\rm B}
\sqrt{{\sf F}_{\rm B}Q_{\rm B}^2
+ {\dot {\sf R}}_{\rm B}^2}
+ \case{1}{2} {\sf R}_{\rm B} {\dot {\sf R}}_{\rm B}
\ln \left(
{\sqrt{{\sf F}_{\rm B}Q_{\rm B}^2
+ {\dot {\sf R}}_{\rm B}^2} - {\dot {\sf R}}_{\rm B}
\over
\sqrt{{\sf F}_{\rm B}Q_{\rm B}^2
+ {\dot {\sf R}}_{\rm B}^2} + {\dot {\sf R}}_{\rm B}
}
\right)
\ \ .
\label{app:bfhB}
\end{eqnarray}
\end{mathletters}%
Here ${\sf R}_{\rm B}$ ( ${\sf R}_{\rm S}$) and $Q_{\rm B}^2$ 
($Q_{\rm S}^2$) are the values of ${\sf R}$
and $Q^2$ at the timelike (spacelike) boundary $r=1$ ($r=0$), 
and ${\sf F} = 1 -
2 {\bf m} {\sf R}^{-1}$. ${\sf R}_{\rm B}$, $Q_{\rm B}^2$,
${\sf R}_{\rm S}$, and $Q_{\rm S}^2$
are considered to be prescribed functions of time,
satisfying ${\sf R}_{\rm B, S}>0$ and $Q_{\rm B, S}^2>0$.
Note that ${\bf H}$ is, in general, explicitly time-dependent.

The variational principle associated with the reduced
action~(\ref{app:S-red}) fixes the initial and final values of~${\bf m}$.
The equations of motion are
\begin{mathletters}
\label{app:red-eom}
\begin{eqnarray}
{\dot {\bf m}} &=& 0
\ \ ,
\label{app:red-eom1}
\\
\noalign{\smallskip}
{\dot {\bf p}} &=&
- {\partial {\bf H} \over \partial {\bf m}}
\nonumber
\\
&=&
\sqrt{{\sf F}_{\rm S}Q_{\rm S}^2
+ {\dot {\sf R}}_{\rm S}^2}
- \sqrt{{\sf F}_{\rm B}Q_{\rm B}^2
+ {\dot {\sf R}}_{\rm B}^2}
\ \ .
\label{app:red-eom2}
\end{eqnarray}
\end{mathletters}%
The interpretation of (\ref{app:red-eom1}) remains unchanged. To interpret
equation~(\ref{app:red-eom2}), note that ${\bf p}$
equals by (\ref{bfp}) the difference of the Killing times
at the left and right ends of the constant $t$ surface. As the
constant $t$ surface evolves in the Schwarzschild spacetime, the
first term in (\ref{app:red-eom2}) gives the evolution rate of the
Killing time at the left end of the hypersurface, where the hypersurface
terminates at a spacelike surface located completely in the future
dynamical region (see Fig. 4). The second term in
(\ref{app:red-eom2}) gives the negative of the evolution rate of the
Killing time at the right end of the surface, where the surface
terminates at the timelike boundary. The two terms are
generated respectively by ${\bf H}_{\rm S}$~(\ref{app:bfhS}) and ${\bf
H}_{\rm B}$~(\ref{app:bfhB}).

The case of interest is when the `inner' spacelike boundary lies on the
Schwarzschild singularity, i.e., ${\sf R}=0=\dot {\sf R}$, and when
the radius of the `outer' boundary two-sphere does not change in time, 
${\dot {\sf R}}_{\rm B}=0$. In that case ${\bf H}_{\rm S}$ 
(\ref{app:bfhS}) and the second term in ${\bf H}_{\rm B}$
(\ref{app:bfhB}) vanish. One can also make the first term in $\dot{\bf p}$ 
(\ref{app:red-eom2}) vanish provided one restricts the slices to approach 
the surface at $r=0$ in such a way that $Q_S$ vanishes faster than 
$\sqrt{F_S}$ there. The second term in (\ref{app:red-eom2})
is readily understood in terms of the Killing time of a static
Schwarzschild observer, expressed as a function of the proper time
$\int^t dt' \sqrt{Q_{\rm B}^2(t')}$ and the blueshift factor ${\sf
F}_{\rm B}^{-1/2}$. The reduced Hamiltonian is given by
\begin{equation}
{\bf H}=
- B\sqrt{{\sf F}_{\rm B}Q_{\rm B}^2},
\end{equation}
where $B$ is the time-independent value of ${\sf R}_{\rm B}$.
Following the same arguments as given in Sec. \ref{subsec:reduction}, 
we find that the appropriate Hamiltonian under the new boundary conditions 
of this appendix is 
\begin{equation}
\label{app:quasiH}
H= \left(1- \sqrt{1-2{\bf m}B^{-1}}\right)B
\ \ ,
\end{equation}
which is the BY Hamiltonian. Similarly, from the time-reversal symmetry 
of the Kruskal extension of the Schwarzschild spacetime, the BY Hamiltonian 
could also be
interpreted to evolve spatial slices that extend from the box upto 
an inner boundary that is the past white hole spacelike singularity.

\vfil
\pagebreak

\centerline{FIGURE CAPTIONS}
\vskip 0.2in

Figure 1: A bounded spacetime region with boundary consisting of initial
and final spatial hypersurfaces $t=t_1$ and $t=t_2$ and a timelike 
three-surface ${}^3 {\sf B}$. ${}^3 {\sf B}$ itself is the time-evolution
of the two-surface ${\sf B}$ that is the boundary of an arbitrary spatial
slice $\Sigma$.

\vskip 0.1in

Figure 2: The Louko-Whiting choice of a foliation of the Schwarzschild 
spacetime. The spatial slices of this foliation extend from the bifurcation
two-sphere to the box. The initial and final spatial hypersurfaces have
label time $t_1$ and $t_2$, respectively.

\vskip 0.1in

Figure 3: A choice of foliating the Schwarzschild spacetime that is different from
the Louko-Whiting choice. Here the spatial slices extend from the box to
a timelike inner boundary that is located completely inside the hole.

\vskip 0.1in

Figure 4: A second way of foliating the Schwarzschild spacetime that is 
different from the Louko-Whiting choice. Here the inner boundary is the future
spacelike black hole singularity.

\vfil
\pagebreak

\begin{figure}[tb]
\begin{center}\leavevmode\epsfbox{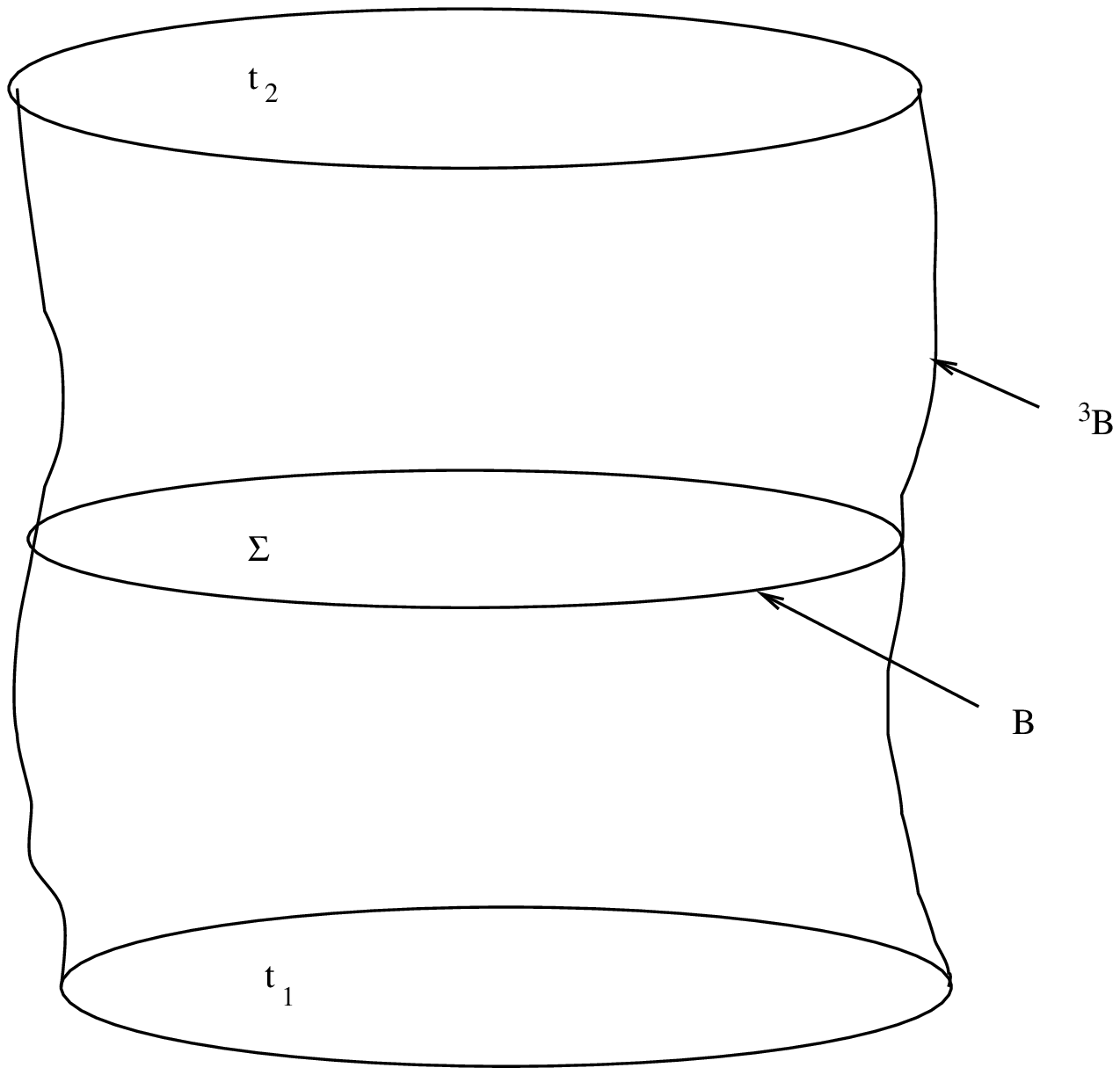}\end{center}
\caption{}
\label{fig1}
\end{figure}

\begin{figure}[tb]
\begin{center}\leavevmode\epsfbox{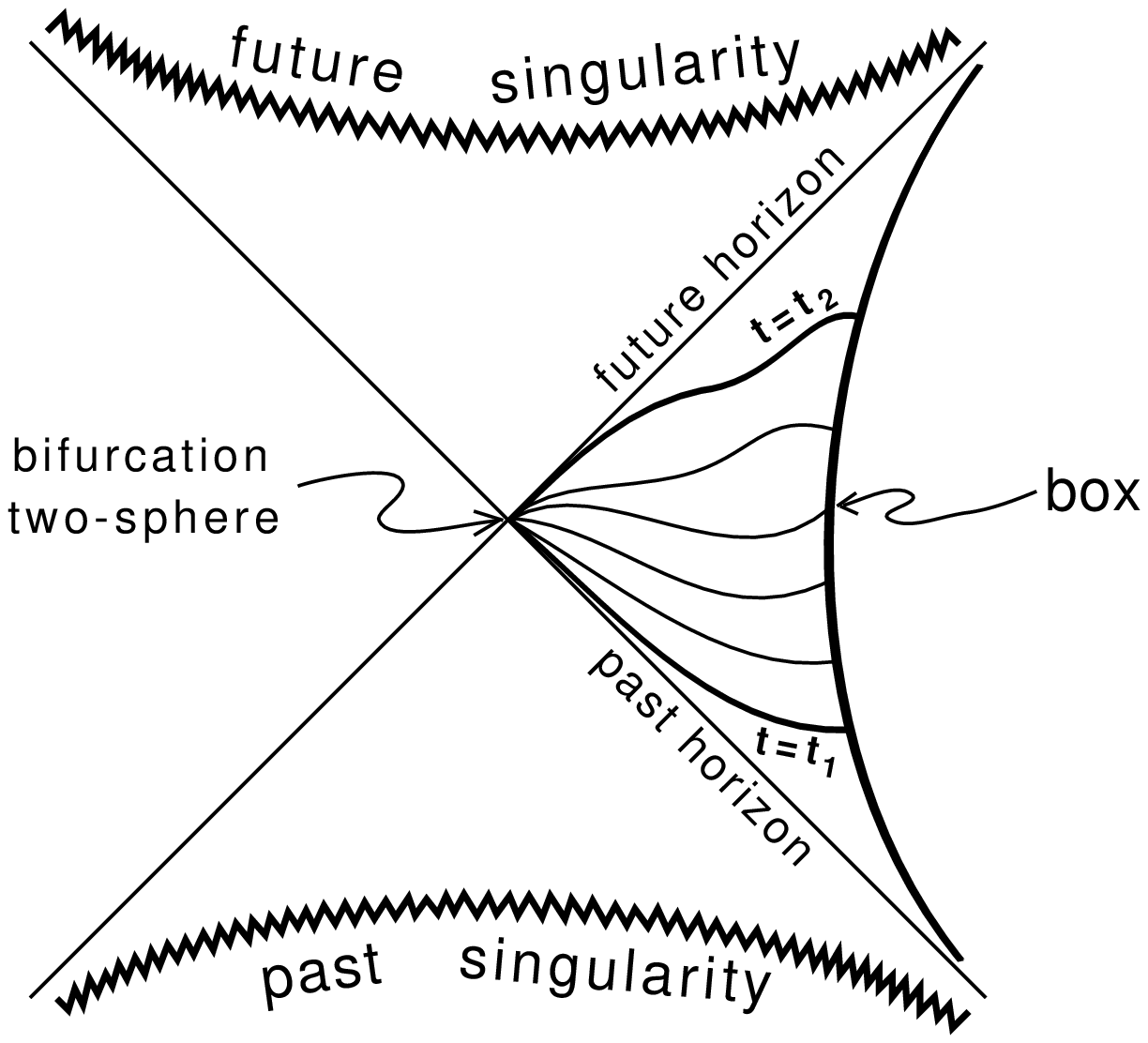}\end{center}
\caption{}
\label{fig2}
\end{figure}

\begin{figure}[tb]
\begin{center}\leavevmode\epsfbox{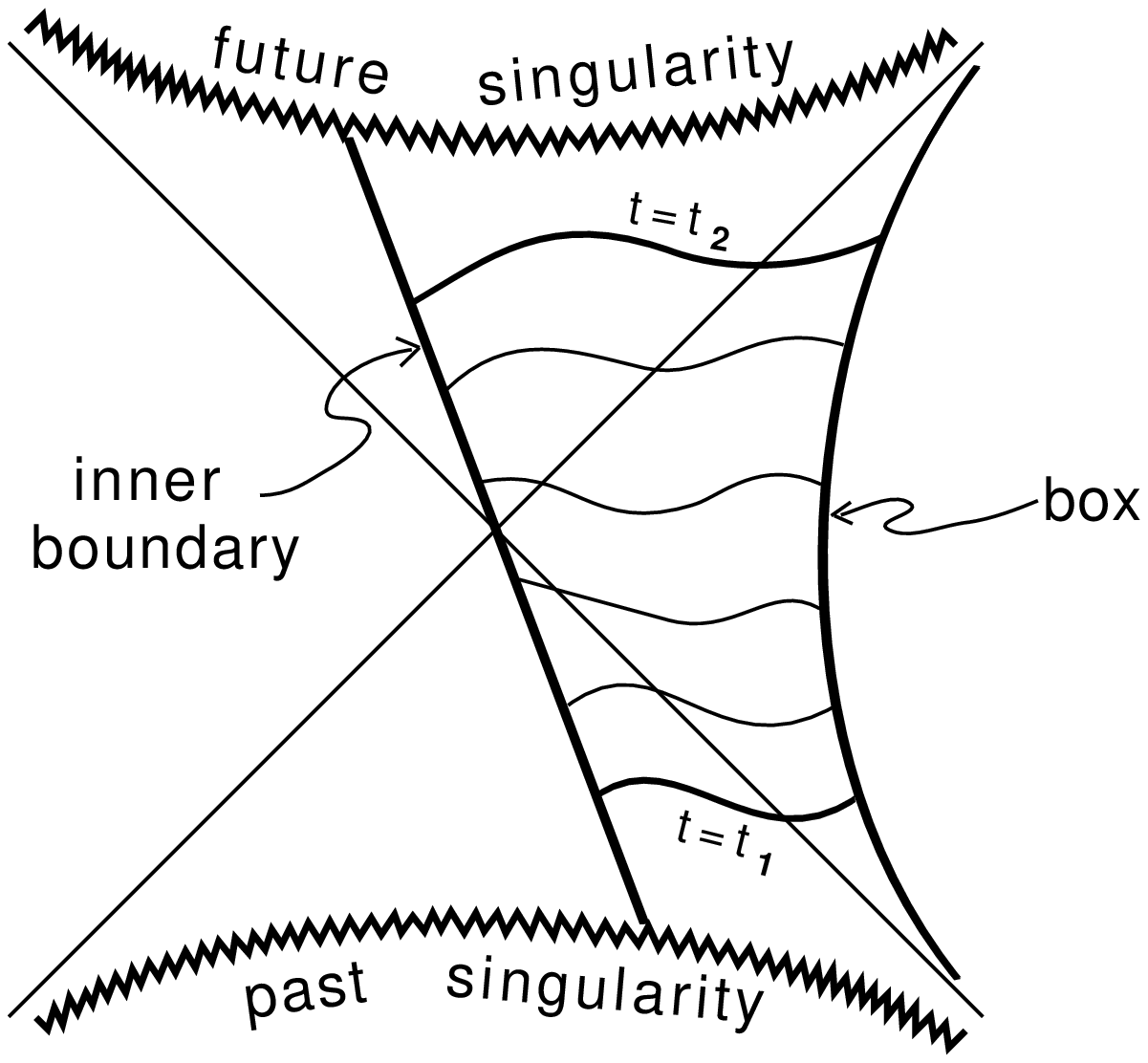}\end{center}
\caption{}
\label{fig3}
\end{figure}

\begin{figure}[tb]
\begin{center}\leavevmode\epsfbox{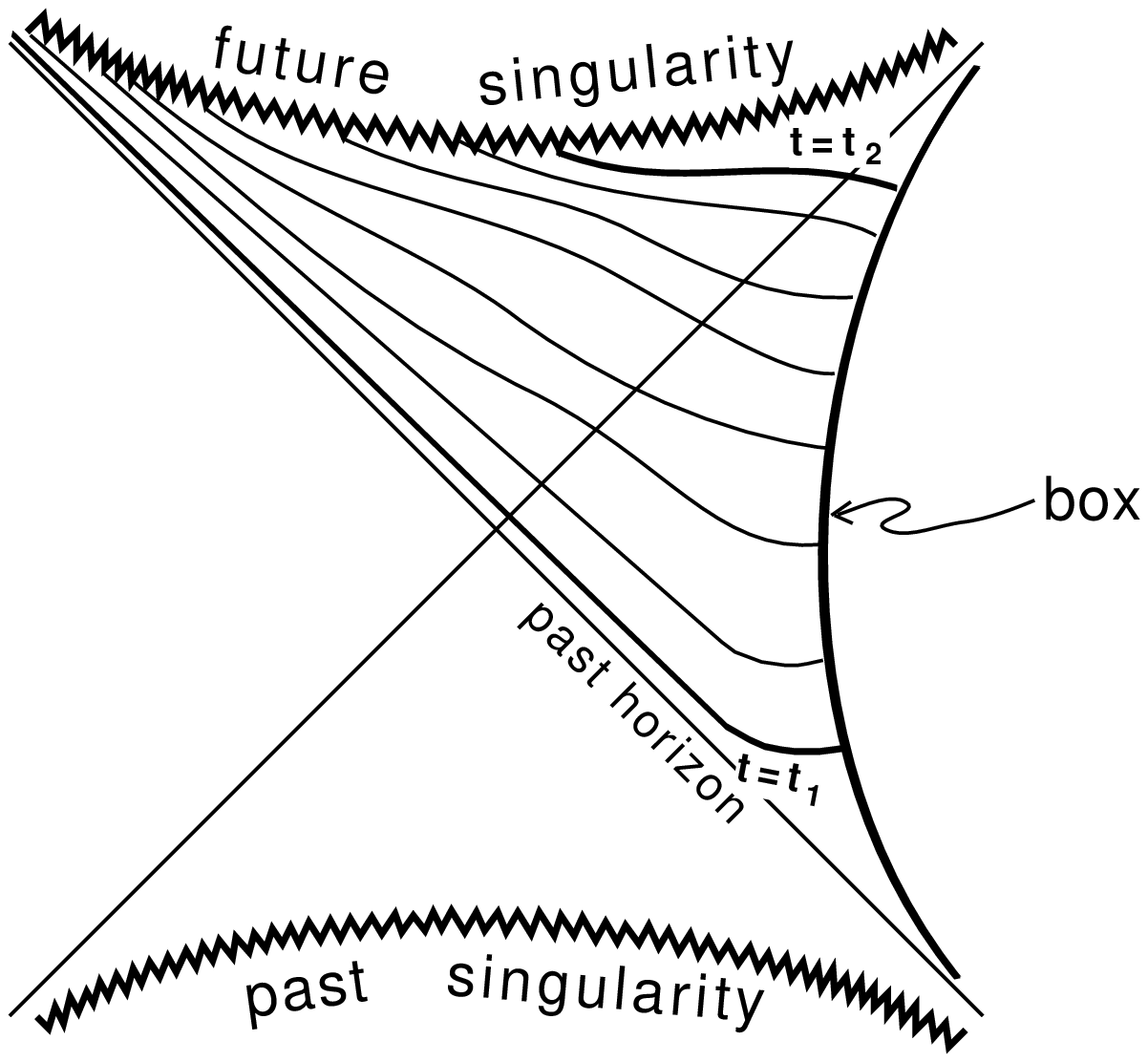}\end{center}
\caption{}
\label{fig4}
\end{figure}


\begin{references}

\bibitem{Hawking1}
S.~W. Hawking, 
Commun.\ Math.\ Phys.\ {\bf 43}, 199 (1975).

\bibitem{Carter}
J.~M. Bardeen, B. Carter and S.~W. Hawking, 
Commun.\ Math.\ Phys.\ {\bf 31}, 161 (1973).  

\bibitem{Bekenstein}
J.~D. Bekenstein, 
Nuovo Cimento Lett. {\bf 4}, 737 (1972); 
Phys. Rev. D {\bf 9}, 3292 (1974). 

\bibitem{Birrell}
N.~D.~Birrell and P.~C.~W.~Davies, {\it Quantum Fields in Curved Space},
(Cambridge University Press, Cambridge, 1982).

\bibitem{BKLS}
L.~Bombelli,
R.~Koul,
J.~Lee, and
R.~D.~Sorkin,
Phys.\ Rev.\ D {\bf 34}, 373 (1986).

\bibitem{FN}
V.~Frolov and
I.~Novikov,
Phys.\ Rev.\ D {\bf48},  4545 (1993).

\bibitem{Srednicki}
M.~Srednicki,
Phys.\ Rev.\ Lett.\ {\bf 71}, 666 (1993);\\
G.~'t Hooft,
Nucl.\ Phys.\ {\bf B256}, 727 (1985);\\
L.~Susskind and
J. Uglum,
Phys.\ Rev.\ D {\bf50},  2700 (1994);\\
M. Maggiore,
Nucl.\ Phys.\ {\bf B429}, 205 (1994);\\
C. Callan and
F. Wilczek,
Phys.\ Lett.\ {\bf B333}, 55 (1994);\\
S. Carlip and
C. Teitelboim, 
Phys.\ Rev.\ D {\bf51}, 622 (1995);\\
S. Carlip,
Phys.\ Rev.\ D {\bf51}, 632 (1995);\\
Y. Peleg,
``Quantum Dust Black Holes'', Brandeis University report no.\
BRX-TH-350, hep-th/93077057 (1993).

\bibitem{GH1}
G.~W. Gibbons and
S.~W. Hawking,
Phys.\ Rev.\ D {\bf15}, 2752 (1977).

\bibitem{hawkingCC}
S.~W. Hawking, in
{\it General Relativity: An Einstein
Centenary Survey,}
edited by S.~W. Hawking and W.~Israel
(Cambridge University Press, Cambridge, 1979).

\bibitem{york1}
J.~W. York,
Phys.\ Rev.\ D {\bf 33}, 2092 (1986).

\bibitem{WYprl}
B.~F. Whiting and
J.~W. York,
 Phys.\ Rev.\ Lett.\ {\bf 61}, 1336 (1988).

\bibitem{whitingCQG}
B.~F. Whiting,
Class.\ Quantum Grav.\ {\bf 7}, 15 (1990).

\bibitem{pagerev}
D.~N. Page, in
{\it Black Hole Physics,}
edited by V.~D. Sabbata and Z.~Zhang
(Kluwer Academic Publishers, Dordrecht, 1992).

\bibitem{LWo}
J.~Louko and
B.~F. Whiting,
Class.\ Quantum Grav.\ {\bf 9},
457 (1992).

\bibitem{BY-microcan}
J.~D. Brown and
J.~W. York,
Phys.\ Rev.\ D {\bf 47},
1420 (1993).

\bibitem{MW}
J.~Melmed and
B.~F. Whiting,
Phys.\ Rev.\ D {\bf 49}, 907
(1994).

\bibitem{TK}
T.~Thiemann and 
H.~A.~Kastrup, Nucl.\ Phys.\ {\bf B399}, 211 (1993);
H.~A.~Kastrup and
T.~Thiemann, Nucl.\ Phys.\ {\bf B425}, 665 (1994);
T.~Thiemann, Int.\ J.\ Mod.\ Phys.\ D {\bf 3}, 293 (1994).

\bibitem{KVK}
K.~V. Kucha\v{r},
Phys.\ Rev.\ D {\bf 50}, 3961
(1994).

\bibitem{LW}
J. Louko and
B.~F. Whiting,
Phys.\ Rev.\ D {\bf 51}, 5583
(1995).

\bibitem{BY-quasilocal}
J.~D. Brown and
J.~W. York,
Phys.\ Rev.\ D {\bf 47}, 1407 (1993).

\bibitem{teitel-dirac}
C.~Teitelboim,
Ann.\ Phys.\ (NY) {\bf 79}, 524 (1973).

\bibitem{Witten}
E.~Witten,
Phys.\ Rev.\ D {\bf 44}, 314 (1991).

\bibitem{BLPP}
S.~Bose,
J.~Louko,
L.~Parker, and
Y.~Peleg,
Phys.\ Rev.\ D {\bf 53}, 7089 (1996).

\end{references}
\end{document}